# Using Interpretable Machine Learning to Massively Increase the Number of Antibody-Virus Interactions Across Studies

**Authors:** Tal Einav[1]*, Rong Ma[2]*

**Affiliations:**

[1]Basic Sciences Division and Computational Biology Program, Fred Hutchinson Cancer Research Center, Seattle, Washington, United States of America

[2]Department of Statistics, Stanford University, Stanford, California, United States of America

*Authors contributed equally to this work

Correspondence should be addressed to teinav@fredhutch.org

**Abstract**

A central challenge in every field of biology is to use existing measurements to predict the outcomes of future experiments. In this work, we consider the wealth of antibody inhibition data against variants of the influenza virus. Due to this virus's genetic diversity and evolvability, the variants examined in one study will often have little-to-no overlap with other studies, making it difficult to discern common patterns or unify datasets for further analysis. To that end, we develop a computational framework that predicts how an antibody or serum would inhibit any variant from *any other study*. We use this framework to greatly expand seven influenza datasets utilizing hemagglutination inhibition, validating our method upon 200,000 existing measurements and predicting 2,000,000 new values±uncertainties. With these new values, we quantify the transferability between seven vaccination and infection studies in humans and ferrets, show that the serum potency is negatively correlated with breadth, and present a tool for pandemic preparedness. This data-driven approach does not require any information beyond each virus's name and measurements, and even datasets with as few as 5 viruses can be expanded, making this approach widely applicable. Future influenza studies using hemagglutination inhibition can directly utilize our curated datasets to predict newly measured antibody responses against ≈80 H3N2 influenza viruses from 1968-2011, whereas immunological studies utilizing other viruses or a different assay only need a single partially-overlapping dataset to extend their work. In essence, this approach enables a shift in perspective when analyzing data from "what you see is what you get" into "what anyone sees is what everyone gets."

**Introduction**

Our understanding of how antibody-mediated immunity drives viral evolution and escape relies upon painstaking measurements of antibody binding, inhibition, or neutralization against variants of concern (Petrova and Russell, 2017). Every interaction is unique because: (1) the antibody response (serum) changes even in the absence of viral exposure and (2) for rapidly evolving viruses such as influenza, the specific variants examined in one study will often have little-to-no overlap with other studies (Figure 1). This lack of cross-talk hampers our ability to comprehensively characterize viral antigenicity, predict the outcomes of viral evolution, and make critical decisions such as whether to update the annual influenza vaccine (Morris et al., 2018).

In this work, we develop a new cross-study matrix completion algorithm that leverages patterns in antibody-virus inhibition data to infer unmeasured interactions. Specifically, we demonstrate that multiple datasets can be combined to predict the behavior of viruses that were entirely absent from one or more datasets (*e.g.*, Figure 2A, predicting values for the green viruses in Dataset 2 and the gray viruses in Dataset 1). Whereas past efforts could only predict values for partially-observed viruses within a single dataset (*i.e.*, predicting the red squares for the green/blue viruses in Dataset 1 or the blue/gray viruses in Dataset 2) (Cai et al., 2010; Ndifon, 2011; Einav and Cleary, 2022), here we predict data for viruses that do not yet have a single measurement in a dataset.

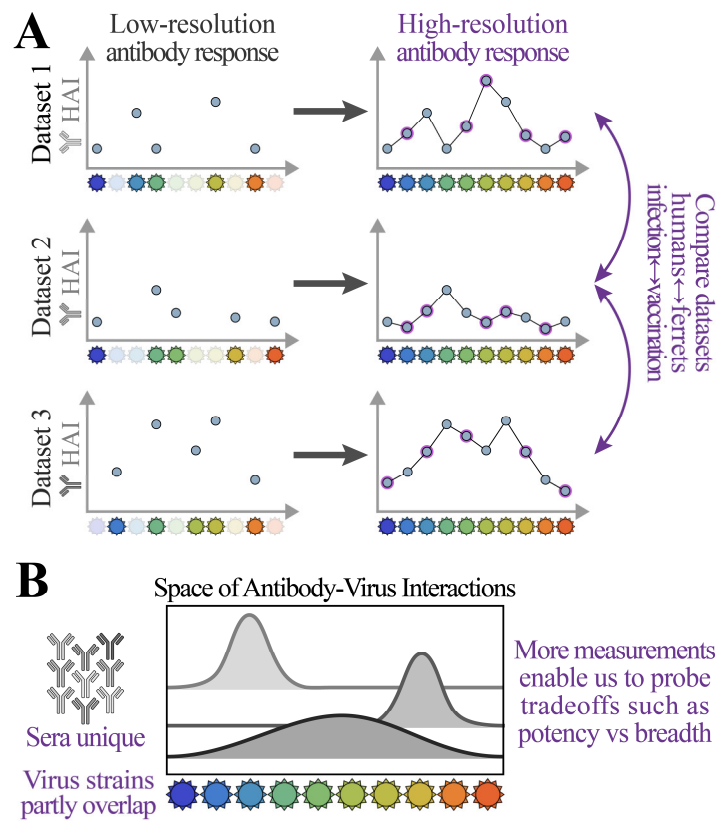

**Figure 1. Challenges of comparing antibody-virus datasets.** (A) We develop a framework that predicts antibody responses (*e.g.*, binding, hemagglutination inhibition [HAI], or neutralization) of any serum against viruses from any other dataset, enabling direct cross-study comparison. (B) Since each serum is unique and virus panels often only partially overlap, these expanded measurements are necessary to characterize the limits of the antibody response or quantify tradeoffs between key features such as potency (the strength of a response) versus breadth (how many viruses are inhibited).

Algorithms that predict the behavior of large virus panels are crucial because they render the immunological landscape in higher resolution, helping to reveal which viruses are potently inhibited and which escape antibody immunity (Morris et al., 2018). For example, polyclonal human sera that strongly neutralize one virus may exhibit 10x weaker neutralization against a virus with one additional mutation (Lee et al., 2019b). Given the immense diversity and rapid evolution of viruses, it behooves us to pool together measurements and build a more comprehensive description of serum behavior.

Even when each dataset is individually complete, many interactions can still be inferred by combining studies. The seven datasets examined in this work measured 60–100% of interactions between their specific virus panel and sera, yet against an expanded virus panel containing every variant from these studies, fewer than 10% of interactions are measured. Moreover, the missing entries are highly structured, with entire columns (representing viruses, Figure 2) missing from each dataset. This introduces unique challenges, since most matrix completion or imputation methods require missing entries to be randomly distributed (Candes and Recht, 2009; Cai et al., 2010; Candes and Tao, 2010; Candes and Plan, 2010; Keshavan et al., 2010; Little and Rubin, 2019), and the few methods tailored for structured missing data focus on special classes of generative models that are less effective in this context (Cai et al., 2016; Xue and Qu, 2021; Xue et al., 2021). In contrast, we construct a framework that harnesses the specific structure of these missing values, enabling us to predict over 2,000,000 new values comprising the remaining 90% of interactions.

The key feature we develop that enables matrix completion across studies is error quantification. Despite numerous algorithms to infer missing values, only a few methods exist that can quantify the error of these predictions when missing values are randomly distributed across a dataset (Carpentier et al., 2018; Chen et al., 2019), and to our knowledge no methods estimate error for general patterns of missing data. Since we do not know *a priori* whether datasets can inform one another, it is crucial to quantify the confidence of these predictions. Our framework uses a data-driven approach to quantify the *individual error* of each prediction, so that users can focus on high-confidence inferences (*e.g.*, those with ≤4-fold error) or search for additional datasets that would further reduce this uncertainty.

Our results provide guiding principles in data acquisition and promote the discovery of new mechanisms in several key ways: (*i*) Existing antibody-virus datasets can be unified to predict each serum against any virus, providing a massive expansion of data and fine-grained resolution into these antibody responses. (*ii*) This expanded virus panel enables an unprecedented direct comparison between human↔ferret and vaccination↔infection studies, quantifying how distinct the antibody responses are in each category. (*iii*) Using the expanded data, we explore the relation between two key features of the antibody response, showing the tradeoff between potency and breadth. (*iv*) We demonstrate an application for pandemic preparedness, where the inhibition of a new variant measured in one study is immediately extrapolated to previous studies. (*v*) We pave the way for future studies to leverage matrix completion, saving time and resources by measuring a substantially smaller subset of viruses. In particular, our approach can determine which subset of viruses will be maximally informative and quantify the benefits of measuring each additional virus.

Although this work focuses on antibody-virus inhibition measurements for influenza, it readily generalizes to other viruses, other assays (*e.g.*, between multiple studies, all using binding or neutralization), and even other applications involving intrinsically low-dimensional datasets.

## Results
**The Low Dimensionality of Antibody-Virus Interactions Empowers Matrix Completion**
Given the vast diversity of antibodies, it is easy to imagine that serum responses cannot inform one another. Indeed, many factors including age, geographic location, frequency/type of vaccinations, and infection history shape the antibody response and influence how it responds to a vaccine or a new viral threat (Kim et al., 2012; Fonville et al., 2014; Thompson et al., 2016; Gouma et al., 2020; Fox et al., 2022).

Yet much of the heterogeneity of antibody responses collapses when we consider their functional behavior such as binding, inhibition, or neutralization against viruses. Previous work has shown that antibody-virus inhibition data are intrinsically low-dimensional (Lapedes and Farber, 2001), which spurred applications ranging from antigenic maps to the recovery of missing values from partially observed data (Smith et al., 2004; Cai et al., 2010; Ndifon, 2011; Einav and Cleary, 2022). However, these efforts have almost exclusively focused on individual datasets of ferret sera generated under controlled laboratory conditions, circumventing the many obstacles of predicting across heterogeneous human studies.

To appreciate the magnitude of this challenge, note that it is only possible to predict measurements for a virus in Dataset 1 (*e.g.*, the virus-of-interest in Figure 2A, boxed in gold) using Dataset 2 provided that: (1) The virus-of-interest is related to at least some of the viruses present in both studies (the other blue viruses), (2) the transferability of information between Dataset 1 and Dataset 2 is quantified, accounting for potential systematic differences in the data, and (3) there are a sufficient number of antibody responses and viruses in Dataset 2 to infer the missing values. If these conditions are met, it intuitively makes sense that the patterns between the blue viruses inferred in Dataset 2 can predict the virus-of-interest in Dataset 1.

In this work, we first demonstrate the accuracy of matrix completion by withholding *all* measurements from one virus in one dataset [Figure 2A, gold boxes] and using the other datasets to generate *predictions±errors*, where each *error* quantifies the uncertainty of a *prediction*. Although we seek accurate predictions with low estimated error, it may be impossible to accurately predict some interactions (*e.g.*, measurements of viruses from 2000-2010 may not be able to predict a distant virus from 1970), and those error estimates should be larger to faithfully reflect this uncertainty.

In the following sections, we develop a matrix completion algorithm that incorporates the features described above. We then analyze seven large serological studies containing hemagglutination inhibition (HAI) measurements for human or animal antibody responses. We first validate our approach by withholding and predicting known data and then apply matrix completion across these studies to greatly extend their measurements.

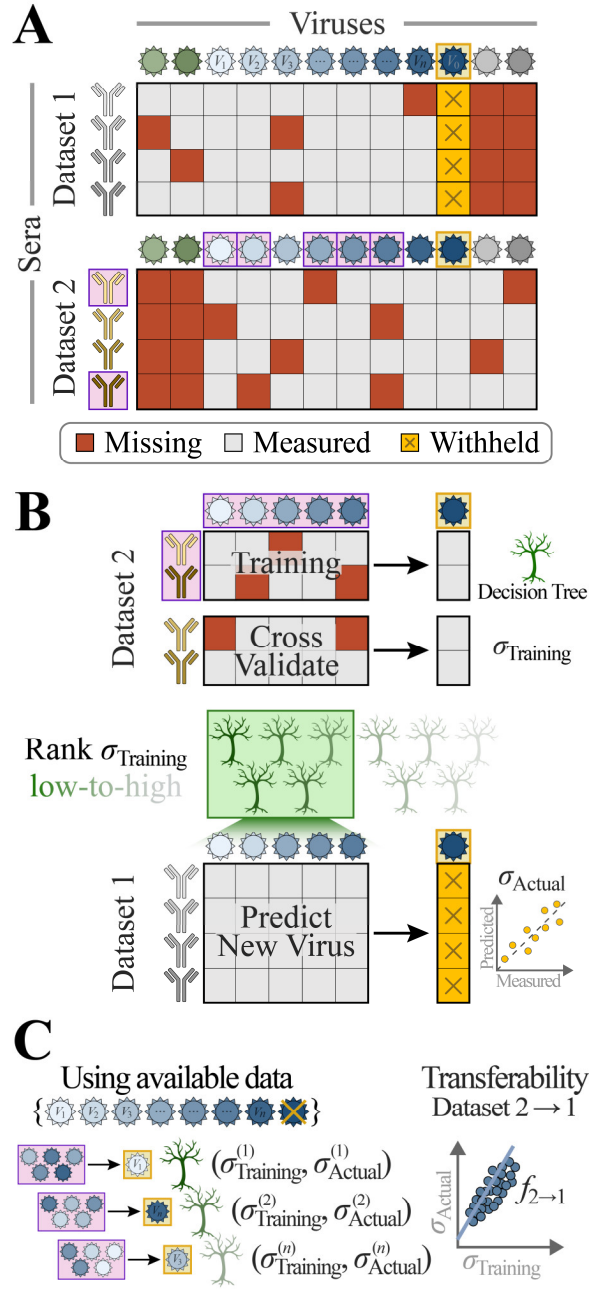

**Figure 2. Combining datasets to predict values and uncertainties for missing viruses.** (A) Schematic of data availability: two studies measure antibody responses against some overlapping viruses (shades of blue) and some unique viruses (green/gray). Studies may have different fractions of missing values (dark-red boxes) and measured values (gray). To test whether virus behavior can be inferred across studies, we predict the titers of a virus in Dataset 1 ($V_0$, gold squares), using measurements from the overlapping viruses ($V_1$-$V_n$) as features in a random forest model. (B) We train a decision tree model using a random subset of antibodies and viruses from Dataset 2 (boxed in purple), cross-validate against the remaining antibody responses in Dataset 2, and compute the root-mean-squared error (RMSE, denoted by $\sigma_{Training}$). Multiple decision trees are trained, and the average from the 5 trees with the lowest error are used as the model going forward. Applying this model to Dataset 1 (which was not used during training) yields the desired predictions, whose RMSE is given by $\sigma_{Actual}$. We repeat this process, withholding each virus in every dataset. (C) To estimate the prediction error $\sigma_{Actual}$ (which we are not allowed to directly compute because $V_0$'s titers are withheld), we define the transferability relation $f_{2\rightarrow 1}$ between the training error $\sigma_{Training}$ in Dataset 2 and actual error $\sigma_{Actual}$ in Dataset 1 using the decision trees that predict viruses $V_1$-$V_n$ (without using $V_0$). We apply this relation to $\sigma_{Training}$ for $V_0$ to estimate its prediction error in Dataset 1.

## Cross-Study Matrix Completion using a Random Forest

We first predict virus behavior between two studies before considering multiple studies. Figure 2 and Algorithm 1 summarize our *leave-one-out* analysis, where a virus-of-interest $V_0$ is withheld from one dataset (Figure 2A, blue virus boxed in gold). We create multiple decision trees using a subset of overlapping viruses $V_1, V_2...V_n$ as features and a subset of antibody responses within Dataset 2 for training (Methods). These trees are cross-validated using the remaining antibody responses from Dataset 2 to quantify each tree's error $\sigma_{Training}$, and we predict the $V_0$ in Dataset 1 using the average of the values±errors from the 5 best trees with the lowest error (Figure 2B).

One potential pitfall of this approach is that the estimated error $\sigma_{Training}$ derived from Dataset 2 will almost always underestimate the true error for these predictions ($\sigma_{Actual}$) in Dataset 1, since the antibody responses in both studies may be very distinct (*e.g.*, responses measured in different years or collected from people/animals with different infection histories). Hence, the relationships between viruses in Dataset 2 may not reflect the relationships between these same viruses in Dataset 1.

To correct for this effect, we quantify the transferability map $f_{2 \to 1}(x)$ between Dataset 2 and Dataset 1, so that for the majority of decision trees, $f_{2 \to 1}(\sigma_{Training}$ from Dataset 2$) \geq \sigma_{Actual}$ in Dataset 1. To that end, we repeat the above algorithm, but rather than inferring values for $V_0$, we now predict each of the overlapping viruses $V_1$-$V_n$ with measurements in *both* datasets (Figure 2C, Algorithm 2). For these viruses, we can directly compare their estimator $\sigma_{Training}$ and true $\sigma_{Actual}$ to infer $f_{2 \to 1}$, which we found was well-characterized by a simple linear relationship [Figure S1; note that $f_{2 \to 1}$ represents an upper bound and not an equality]. Finally, we use this relation to compute the prediction error for $V_0$ in Dataset 1, $\sigma_{Predict} \equiv f_{2 \to 1}(\sigma_{Training})$. In this way, both the values and errors are inferred using a generic, data-driven approach that can be applied to diverse datasets.

## Leave-One-Out: Inferring Virus Behavior without a Single Measurement

To assess matrix completion across studies, we applied it to three increasingly difficult scenarios: (1) between two highly-similar human vaccination studies, (2) between a human infection and human vaccination study, and (3) between a ferret infection and human vaccination study. We expect that prediction accuracy will decrease as the datasets become more distinct, resulting in a larger error ($\sigma_{Actual}$) and larger estimated uncertainty ($\sigma_{Predict}$).

For these predictions, we utilized the Fonville influenza datasets consisting of six studies – four human vaccination studies (Dataset$_{Vac,1-4}$), one human infection study (Dataset$_{Infect,1}$), and one ferret infection study (Dataset$_{Ferret}$) (Fonville et al., 2014). In each study, sera were measured against a panel of H3N2 viruses using hemagglutination inhibition. Collectively, these studies contained 81 viruses, and each virus was measured in at least two studies.

We first predicted values for the virus $V_0$=A/Auckland/5/1996 in the most recent vaccination study (Dataset$_{Vac,4}$) using data from another vaccination study (Dataset$_{Vac,3}$) carried out in the preceding year and in the same geographic location [Table 1]. After training our decision trees, we found that the two studies had near-perfect transferability ($\sigma_{Predict}=f_{Vac,3 \to Vac,4}(\sigma_{Training}) \approx \sigma_{Training}$), suggesting that there is no penalty in extrapolating virus behavior between these datasets. More precisely, if there exist five viruses $V_1$-$V_5$ that can accurately predict $V_0$'s measurements in Dataset$_{Vac,3}$, then $V_1$-$V_5$ will predict $V_0$ equally well in Dataset$_{Vac,4}$.

Indeed, we found multiple such decision trees, with which we predicted $V_0$ together with the uncertainty of $\sigma_{Predict}$=1.8-fold, meaning that each titer $t$ is expected to lie between $t/1.8$ and $t \cdot 1.8$ with at least 68% probability [or equivalently that $\log_{10}(t)$ has a standard deviation of $\log_{10}(1.8)$] (top panel in Figure 3A, gray bands represent $\sigma_{Predict}$). Moreover, this estimated uncertainty closely matched the true error $\sigma_{Actual}$=1.6-fold. To put these results into perspective, the HAI assay has roughly 2-fold error (*i.e.*, repeated measurements differ by 2-fold 50% of the time and by 4-fold 10% of the time, Methods), implying that these predictions are as good as possible given experimental error.

When we inferred every other virus between Datasets$_{Vac,3 \to Vac,4}$, we consistently found the same highly accurate predictions $\sigma_{Predict} \approx \sigma_{Actual} \approx$2-fold [Figure S2A]. As an alternative way of quantifying error, we plotted the distribution of predictions within 0.5, 1.0, 1.5… standard deviations from the measurement, which we compare against a folded Gaussian distribution [Figure 3A, bottom]. For example, 81% of predictions were within one standard deviation, somewhat larger than the 68% expected for a Gaussian, which confirms that the prediction error was slightly overestimated.

We next predicted values for $V_0$=A/Netherlands/620/1989 between a human infection and vaccination study (Dataset$_{Infect,1 \to Vac,4}$). In this case, the predicted values were also highly accurate with true error $\sigma_{Actual}$=2.2-fold [Figure 3B; remaining viruses predicted in Figure S2B]. When quantifying the uncertainty of these predictions, we found far less transferability of virus behavior ($\sigma_{Predict}$=$f_{Infect,1 \to Vac,4}(\sigma_{Training}) \approx 2.8\sigma_{Training}$, where the larger prefactor of 2.8 indicates less transferability, see Methods), and hence we overestimated the prediction error as $\sigma_{Predict}$=4.0-fold. Lastly, when we predicted values for $V_0$=A/Perth/27/2007 between a ferret infection and human vaccination study (Dataset$_{Ferret \to Vac,4}$), our predictions had increased true error $\sigma_{Actual}$=4.6-fold [Figure 3C], and poor transferability again led to a larger estimated prediction error $\sigma_{Predict}$=6.6-fold. Importantly, this overestimation is purposefully built into the transfer functions $f_{X \to Y}$ whenever Datasets $X$ and $Y$ exhibit very noisy or highly disparate behaviors, so that we err on the side of overestimating rather than underestimating uncertainty whenever prediction is intrinsically difficult. In contrast, the transferability $f_{X \to Y}$ is highly accurate between studies where virus behavior can be consistently extrapolated (*e.g.*, between Datasets$_{Vac,1/2}$ or Datasets$_{Vac,3/4}$; see Figure S1). Moreover, as we show in the following section, the estimated values and error become more precise when we use multiple datasets to infer virus behavior.

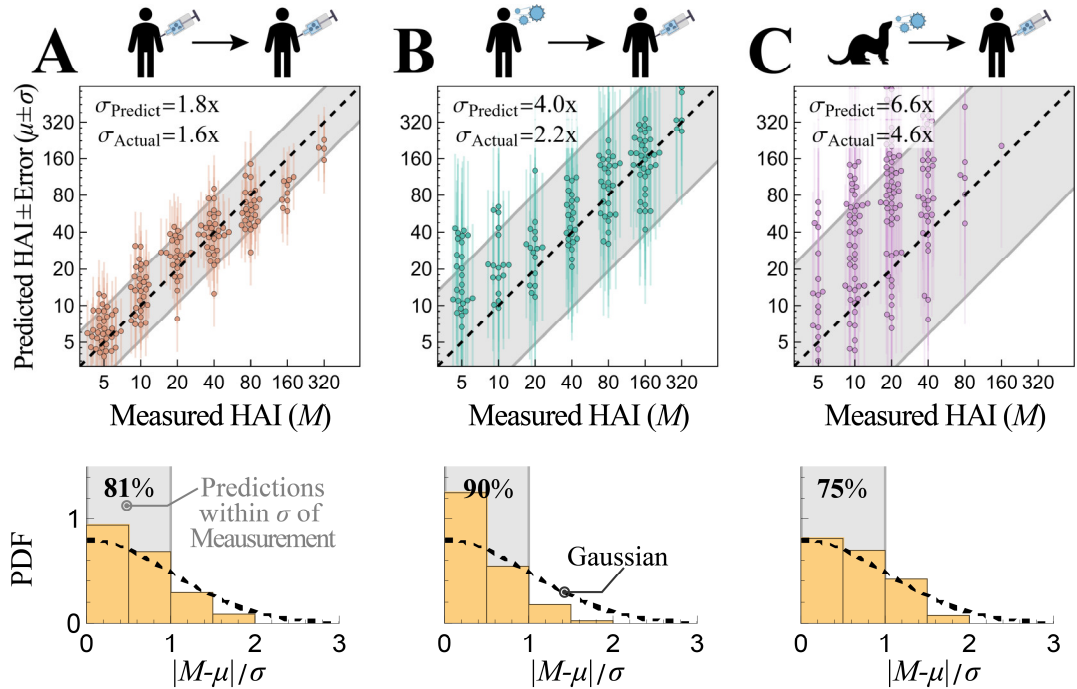

**Figure 3. Predicting virus behavior between two datasets.** Example predictions between two Fonville studies. *Top*, Plots comparing predicted and withheld HAI measurements (which take the discrete values 5, 10, 20…). Estimated error is shown in two ways: (*i*) as vertical lines emanating from each point and (*ii*) by the diagonal gray bands showing $\sigma_{\text{Predict}}$. *Bottom*, Histograms of the standardized absolute prediction errors compared to a standard folded Gaussian distribution [black dashed line]. The fraction of predictions within $1.0\sigma$ are shown in the top left, which can be compared with the expected 68% for the standard folded Gaussian distribution. (A) Predicting A/Auckland/5/1996 between two human vaccination studies (Datasets$_{\text{Vac,3}\to\text{Vac,4}}$). (B) Predicting A/Netherlands/620/1989 between a human infection and human vaccination study (Datasets$_{\text{Infect,1}\to\text{Vac,4}}$). (C) Predicting A/Perth/27/2007 between a ferret infection and human vaccination study (Datasets$_{\text{Ferret}\to\text{Vac,4}}$).

## Combining Influenza Datasets to Predict 200,000 Measurements with ≤3-fold Error

When multiple datasets are available to predict virus behavior in Dataset 1, we obtain *predictions±errors* ($\mu_j \pm \sigma_j$) from Dataset 2→1, Dataset 3→1, Dataset 4→1… These predictions and their errors are combined using the standard Bayesian approach yielding

$$\frac{\Sigma_j (\mu_j/\sigma_j^2)}{\Sigma_j (1/\sigma_j^2)} \pm \frac{1}{[\Sigma_j (1/\sigma_j^2)]^{1/2}}. \qquad \text{(Equation 1)}$$

The uncertainty term in this combined prediction has two key features. First, adding an additional dataset (with predictions $\mu_k \pm \sigma_k$) only decreases the uncertainty. Second, if a highly uninformative dataset is added (with $\sigma_k \to \infty$), it will negligibly affect the cumulative prediction. Therefore, as long as the uncertainty estimates are reasonably precise, datasets do not need to be prescreened before matrix completion, and adding more datasets will always result in lower uncertainty.

To test the accuracy of combining multiple datasets, we performed leave-one-out analysis using all six Fonville studies, systematically withholding every virus in each dataset (311 virus-dataset pairs) and predicting the withheld values using all remaining data. Each dataset measured 35-300 sera against 20-75 viruses (with 81 unique viruses across all six studies) and had 0.5-40% missing values [Figure 4A].

Across all datasets, we predicted these 50,000 measurements with a low error of $\sigma_{Actual}$=2.1-fold (between the measured value and the left-hand side of Equation 1). Upon stratifying these predictions by dataset, we found that the four human vaccination studies were predicted with the highest accuracy (Datasets$_{Vac,1-4}$, $\sigma_{Actual}$≈2-fold) while the human infection study had slightly worse accuracy (Dataset$_{Infect,1}$, $\sigma_{Actual}$=2.6-fold) [Figure 4A]. Remarkably, even the least accurate human→ferret predictions had ≤4-fold error on average ($\sigma_{Actual}$=3.7-fold), demonstrating the potential for these cross-study inferences.

In addition to accurately predicting these values, the estimated error closely matched the true error in every human study ($\sigma_{Predict} \approx \sigma_{Actual}$, Datasets$_{Vac,1-4}$ and Dataset$_{Infect,1}$). In contrast, the uncertainty of the ferret predictions was noticeably overestimated ($\sigma_{Predict}$=6.4-fold, Dataset$_{Ferret}$) because none of the other datasets showed appreciable transferability to the ferret data. Said another way, some viruses behaved similarly in the human and ferret datasets while other viruses did not, and in this highly-variable context our framework may overestimate $\sigma_{Predict}$. Mathematically, a large $\sigma_{Predict}$ arises because training error correlates poorly with prediction error, which results in steep transferability functions for the ferret data [Figure S1].

We visualize the transferability between datasets using a chord diagram [Figure 4B], where wider bands connecting Datasets$_{X \leftrightarrow Y}$ represent larger transferability [Figure S3, Methods]. As expected, there was high transferability between the human vaccine studies carried out in consecutive years (Datasets$_{Vac,1 \leftrightarrow Vac,2}$ and Datasets$_{Vac,3 \leftrightarrow Vac,4}$, Table 1), but far less transferability across vaccine studies more than 10 years apart (Datasets$_{Vac,1 \leftrightarrow Vac,3}$, Datasets$_{Vac,1 \leftrightarrow Vac,4}$, Datasets$_{Vac,2 \leftrightarrow Vac,3}$, or Datasets$_{Vac,2 \leftrightarrow Vac,4}$).

Transferability is not necessarily symmetric, since virus inhibition in Dataset *X* could exhibit all patterns in Dataset *Y* (leading to high transferability from *X*→*Y*) along with unique patterns not seen in Dataset *Y* (resulting in low transferability from *Y*→*X*). For example, all human datasets displayed small transferability to the ferret data, although the ferret dataset could somewhat predict Dataset$_{Infect,1}$ and Datasets$_{Vac,3/4}$; this suggests the ferret responses shows some patterns present in the human data but also display unique phenotypes. As another example, the human infection study carried out from 2007-2012 had high transferability from the human vaccine studies conducted in 2009 and 2010 (Dataset$_{Vac,3/4 \rightarrow Infect,1}$) but showed smaller transferability in the reverse direction.

While these results lay the foundation to compare different datasets and quantify the impact of age, geographic location, and other features on the antibody response, they are not exhaustive characterizations – for example, additional human datasets using other viruses or sera may be able to predict these ferret responses more accurately. The strength of this approach lies in the fact that cross-study relationships are learned in a data-driven manner. In particular, as more datasets are added, the number of predictions between datasets increases while the uncertainty of these predictions decreases.

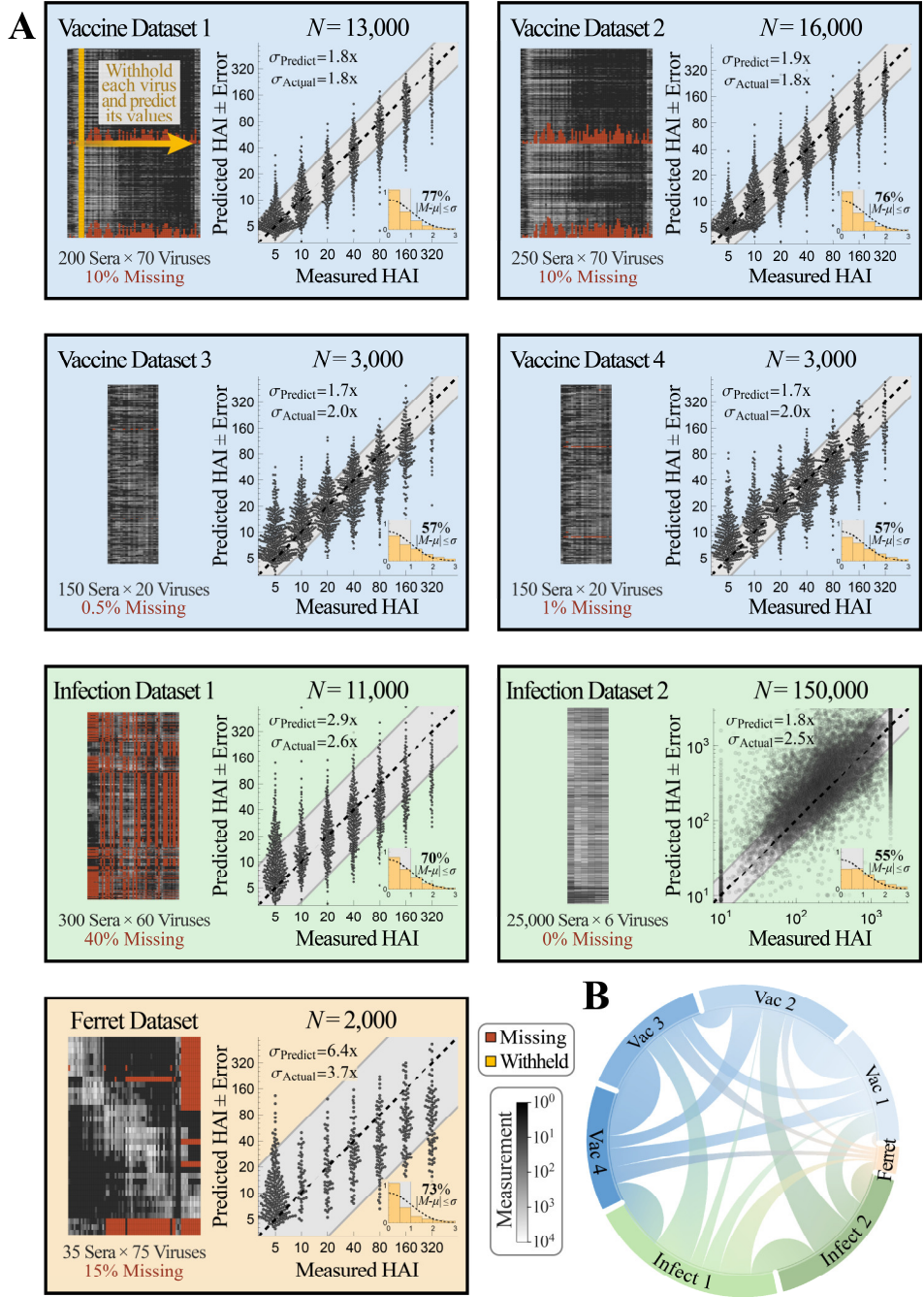

**Figure 4. Validating prediction±error quantification across 200,000 measurements.** (A) We combined seven influenza datasets spanning human vaccination studies (blue boxes), human infection studies (green), and a ferret infection study (orange). Each virus in every dataset was withheld and predicted using the remaining data (shown schematically in gold within the top-left box). We display each dataset (*left*, missing values in dark-red and measurements in grayscale) and the collective predictions for all viruses in that dataset (*right*, gray diagonal bands show the average predicted error $\sigma_{Predict}$). The total number of predictions $N$ from each dataset is shown above its scatterplot; when this number of points is too great to show, we subsampled each distribution evenly while maintaining its shape. The inset at the bottom-right of each plot shows the PDF histogram of error measurements [*y*-axis] that were within 0.5$\sigma$, 1.0$\sigma$, 1.5$\sigma$… [*x*-axis], compared to a standard folded Gaussian distribution (black curve). The fraction of predictions within 1.0$\sigma$ is explicitly written, which can be compared with the expected 68% for a standard folded Gaussian. (B) Chord diagram representing the transferability between datasets. For each arc connecting datasets *X* and *Y*, a larger width at the outer circle represents greater transferability [see Figure S3 and Methods].

**Versatility of Matrix Completion: Predicting Values from a Distinct Assay using only 5 Overlapping Viruses**

To test the limits of our approach, we used the Fonville datasets to predict values from a large-scale serological dataset by Vinh *et al*. where *only 6* influenza viruses were measured against 25,000 sera (Vinh et al., 2021). This exceptionally long-and-skinny matrix is challenging for several reasons. First, after entirely withholding a virus, only 5 other viruses remain to infer its behavior. Furthermore, only 4/6 of the Vinh viruses had exact matches in the Fonville dataset; to utilize the remaining 2 viruses, we associated them with the closest Fonville virus based on their hemagglutinin sequences [Methods].

Second, the Vinh study used protein microarrays to measure serum binding to the HA1 subunit that forms the hemagglutinin head domain. While HAI also measures how antibodies bind to this head domain, such differences in the experimental assay could lead to fundamentally different patterns of virus inhibition, which will result in low transferability and high estimated error. (We also note one superficial difference: the Vinh data span a continuum of values while the Fonville data take on discrete 2-fold increments, but this feature does not affect our algorithm.)

Third, there were only 1,200 sera across all Fonville datasets, and hence predicting the behavior of 25,000 Vinh sera will be impossible if they all exhibit distinct phenotypes. Indeed, any such predictions would only be possible if the behavior of the 6 Vinh viruses can be related to the other Fonville viruses, and if this multitude of serum responses is very degenerate.

Finally, even if accurate predictions are possible, it is hard to predict *a priori* which of the Fonville datasets would be most informative. The Vinh human infection study (Dataset$_{Infect,2}$) was carried out in Vietnam from 2009-2015, suggesting that either the Fonville human infection study (Dataset$_{Infect,1}$, carried out in Vietnam during similar years) or vaccine studies (Datasets$_{Vac,3/4}$ from 2009-2010) would be the best candidates [Table 1]. Surprisingly, we found that the greatest contribution came from an earlier Fonville vaccine study (Dataset$_{Vac,2}$ from 1998) as well as the human infection study [Figure 4B]. (Datasets$_{Vac,3/4}$ could not make any predictions because they only contained 1/6 of the Vinh viruses.)

After growing a forest of decision trees to establish the transferability between the Fonville and Vinh datasets [Figure S1], we predicted the 25,000 serum measurements for all 6 Vinh viruses with an average $\sigma_{Actual}$=2.5-fold error, demonstrating that even a small panel containing 5 viruses can be expanded to predict the behavior of additional strains [Figure 4, Dataset$_{Infect,2}$].

Notably, 5/6 of these viruses (which all circulated between 2003-2011) had a very low $\sigma_{Predict} \approx \sigma_{Actual} \approx$ 2-fold error [Figure S4]. The final Vinh virus circulated three decades earlier (in 1968), and its larger prediction error was underestimated ($\sigma_{Actual}$=4.5-fold, $\sigma_{Predict}$=2.7-fold). This inaccurate uncertainty highlights a shortcoming of any matrix completion algorithm, namely, that when a dataset contains one exceptionally distinct column (*i.e.*, one virus circulating 30 years before all other viruses), its values will not be accurately predicted. Nevertheless, we note that even predicting values with 4.5-fold error can identify the weakest and strongest responses given that the measurements span a 200-fold range.

**Leave-Multi-Out: Designing a Minimal Virus Panel that Maximizes the Information Gained per Experiment**

Given the accuracy of leave-one-out analysis and that only 5 viruses are needed to expand a dataset, we reasoned that most studies include many viruses whose behavior could have been inferred. Pushing this to the extreme, we combined the Fonville and Vinh datasets and performed *leave-multi-out analysis*, where multiple viruses were simultaneously withheld and recovered. Future studies seeking to measure any set of viruses $V_1$-$V_n$ can use a similar approach to select the minimal virus panel that could predict their full data.

In the present search, we sought the minimum viruses needed to recover all Fonville and Vinh measurements with ≤4-fold error; we chose this threshold because we can remove dozens of viruses but still clearly distinguish the most versus least inhibited variants. A virus was randomly selected from a dataset and added to the withheld list if its values, and those of all other withheld viruses, could be predicted with $\sigma_{Predict}$≤4-fold (without using $\sigma_{Actual}$ to confirm these predictions, Methods). In this way, 124 viruses were concurrently withheld, representing 25–60% of the virus panels from every dataset or a total of $N$=70,000 measurements [Figure 5A].

Even with this hefty withheld set, the predictions only exhibited slightly larger errors than during leave-one-out analysis ($\sigma_{Actual}$ between 2.1–3.1-fold for the human datasets and $\sigma_{Actual}$ =3.8-fold for the ferret data). This small increase is due to two competing factors. On the one hand, prediction is far harder with fewer viruses. At the same time, our approach specifically withheld the highly-predictable viruses (with $\sigma_{Predict}$≤4-fold). These factors mostly offset one another, so that the 70,000 measurements exhibited the desired $\sigma_{Actual}$≤4-fold.

The transferability between datasets, computed without the withheld viruses, was similar to the transferability between the full datasets [Figure 5B]. Some connections were lost when there were <5 overlapping viruses between datasets, while other connections were strengthened when the patterns in the remaining data became more similar across studies. Notably, the ferret data now showed some transferability from vaccination Datasets$_{Vac,1/2}$, which resulted in smaller estimated error than in our leave-one-out analysis ($\sigma_{Predict}$=2.9-fold). This emphasizes that transferability depends upon the specific viruses and sera examined, and that some parts of the Fonville human dataset can inform the ferret data. While this uncertainty underestimated the true error of the ferret predictions ($\sigma_{Actual}$=3.8-fold), they remained within the desired 4-fold error threshold. Moreover, in all six human datasets, the estimated uncertainty $\sigma_{Predict}$ closely matched the true error $\sigma_{Actual}$, demonstrating that there is significant potential to predict the behavior of many new viruses within each dataset.

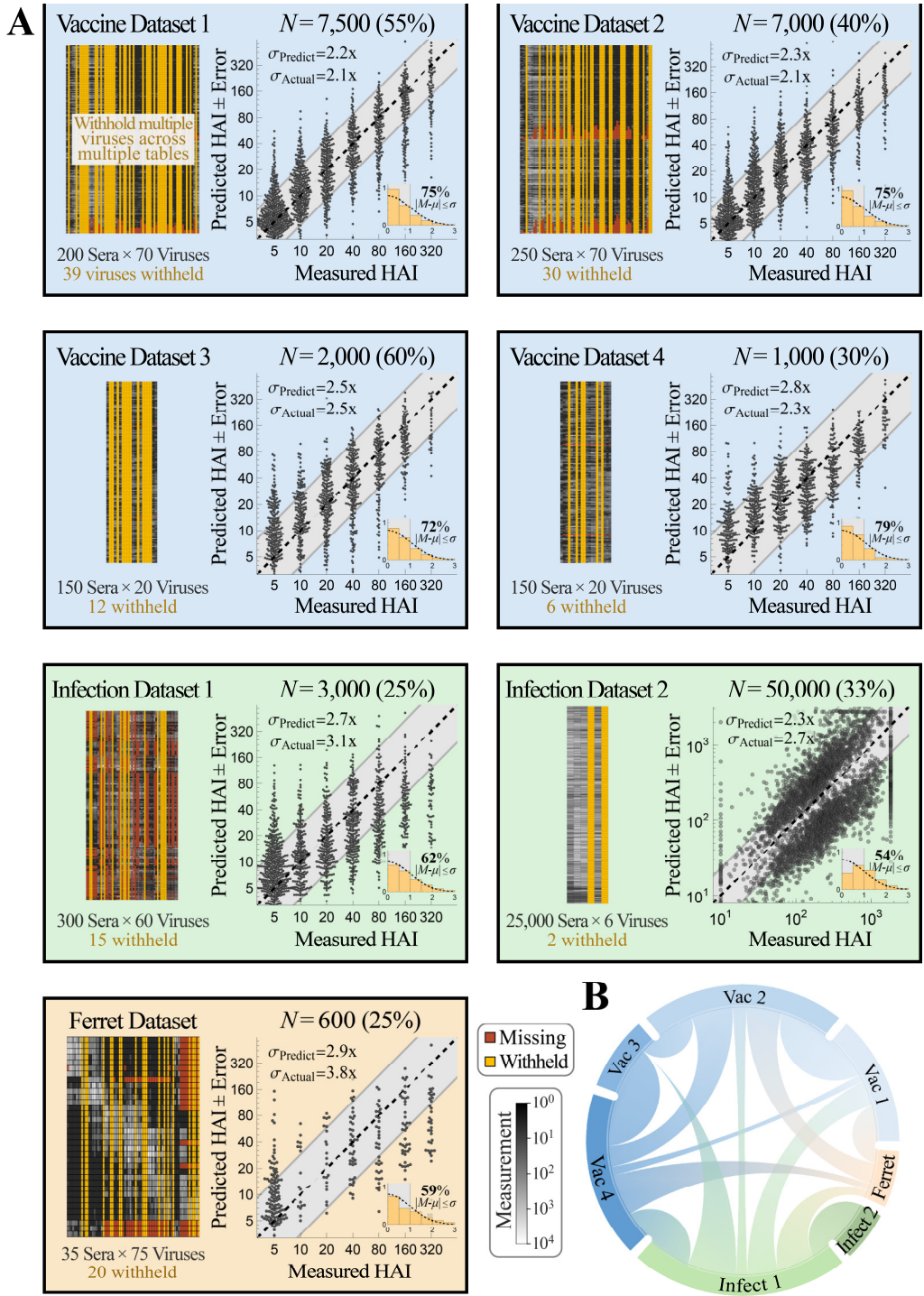

**Figure 5. Simultaneously predicting 124 viruses withheld from multiple datasets.** (A) Viruses were concurrently withheld from each dataset (*left*, gold columns), and their 70,000 values were predicted using the remaining data. We sought to withhold as many viruses as possible while still retaining a low estimated error ($\sigma_{Predict} \leq$ 4-fold), and indeed, the actual prediction error was smaller than 4-fold in every dataset. As in Figure 4, plots and histograms show the collective predictions and error distributions. The plot label enumerates the number of concurrent predictions (and percent of data predicted). (B) Chord diagram representing the transferability between datasets after withholding the viruses. For each arc connecting datasets *X* and *Y*, a larger width at the outer circle represents greater transferability [see Figure S3 and Methods].

## Expanding Datasets with 2·10⁶ New Measurements reveals a Tradeoff between Serum Potency and Breadth

In the previous section, we combined datasets to predict the behavior of additional viruses, validating our approach on 200,000 existing measurements. Future studies can immediately leverage the Fonville datasets to expedite their efforts. If a new dataset contains at least 5 Fonville viruses [green arrows/boxes in Figure 6A], the values±errors for the remaining Fonville viruses can be predicted. Viruses with an acceptably low error [purple in Figure 6A] can be added without requiring any additional experiments.

To demonstrate this process, we first focus on the Vinh dataset where expansion will have the largest impact, since the Vinh virus panel is small (6 viruses) but its serum panel is enormous (25,000 sera). By predicting the interactions between these sera and all 81 unique Fonville viruses, we add 2,000,000 new predictions (more than 10x the number of measurements in the original dataset).

For each Fonville virus $V_0$ that was not measured in the Vinh dataset, we grew a forest of decision trees as described above, with the minor modification that the 5 features must be restricted to the Vinh viruses to enable this expansion. The top trees were combined with the transferability functions (shown in Figure S1) to predict the values±errors for $V_0$ (Figure S5).

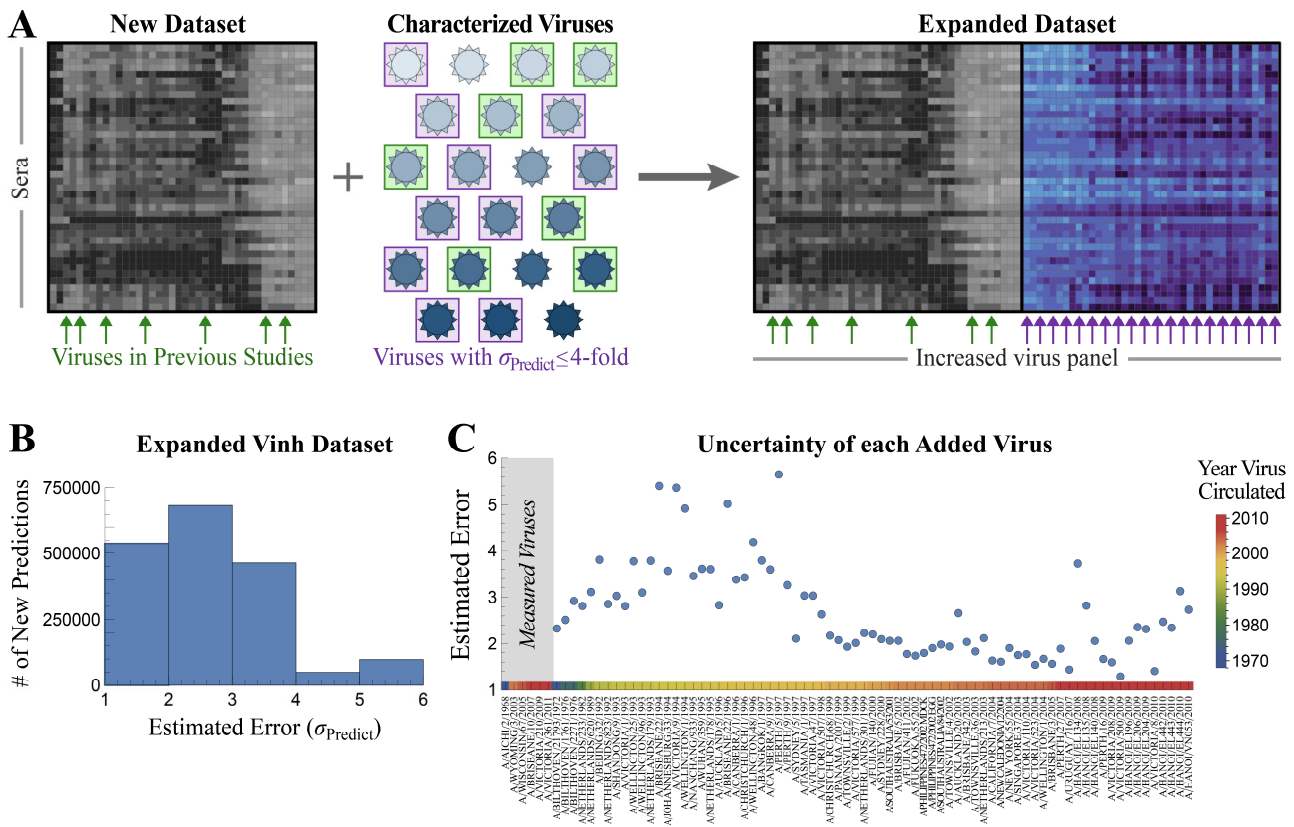

**Figure 6. Expanding the Vinh dataset with 75 additional viruses.** (A) If a new study contains at least 5 previously characterized viruses [green boxes and arrows], we can predict the behavior of all previously characterized viruses in the new dataset. Those with an acceptable error (*e.g.*, ≤4-fold error boxed in purple) are used to expand the dataset. (B) Distribution of the estimated uncertainty $\sigma_{Predict}$ when predicting how each Fonville virus inhibits the 25,000 Vinh sera. Most viruses are estimated with ≤4-fold error. (C) Estimated uncertainty of each virus. The six viruses on the left represent the Vinh virus panel. Colors at the bottom represent the year each virus circulated.

The majority of the added Fonville viruses (68/75) were predicted with a low uncertainty of $\sigma_{Predict} \leq 4$-fold [Figure 6B]. As expected, viruses circulating around the same time as the Vinh panel (1968 or 2003-2011) tended to have the lowest uncertainty, whereas viruses from the 1990s had the largest uncertainty [Figure 6C]. To confirm these estimates, we restricted the Fonville datasets to these same 6 viruses and expanded out, finding that any virus with $\sigma_{Predict} \leq 6$-fold prediction error (as is the case for all Vinh predictions) had a true error $\sigma_{Actual} \leq 6$-fold [Figure S6]. We similarly expanded the Fonville datasets, adding 175 new virus columns across the six studies (Figure S5, extended datasets also provided on GitHub). In addition, dimensionality reduction via UMAP recovered a linear trend from the oldest-to-newest viruses in both the Fonville and Vinh datasets; this trend is especially noteworthy in the latter case, since we did not supply the circulation year for the 75 inferred viruses, yet we can discern its impact on the resulting data (Figure S7).

For each Vinh serum, this expansion fills in the 3.5-decade gap between 1968-2003 by predicting 47 additional viruses, as well as adding another 28 measurements between 2003-2011 (Figure 7A, new interactions highlighted in purple). We also predicted dozens of new viruses in the vaccine studies, and for some sera this increased resolution revealed a more jagged landscape than what was apparent from the direct measurements (Figure 7A). Although HAI titers tend to be similar for viruses circulating around the same time, exceptions do arise (*e.g.*, A/Tasmania/1/1997 vs A/Perth/5/1997 or A/Hanoi/EL201/2009 vs A/Hanoi/EL134/2008 had >4-fold difference in their predicted titers), and our expanded data reveal these functional differences between variants.

The expanded data also enable a direct comparison of sera across studies, something that is exceedingly difficult with the original measurements given that none of the 81 viruses were in all seven datasets (and the different dynamic ranges of the Vinh and Fonville datasets). Figure 7A shows that an antibody response may be potent against older strains circulating before 2000 but weak against newer variants [bottom], highly specific against strains from 1980-2000 with specific vulnerabilities to viruses from 1976 [middle], or relatively uniformly against the entire virus panel [top].

We next used the expanded data to probe a fundamental but often-unappreciated property of the antibody response, namely, the tradeoff between serum potency and breadth. Given a set of viruses circulating within ΔVirus Years of each other [top of Figure 7B shows an example with ΔVirus Years=2], how potently can a serum inhibit all of these viruses simultaneously? For each serum and any set of viruses spanning ΔVirus Years, we computed $HAI_{min}$ (the minimum titer against this set of viruses), and we show the maximum $HAI_{min}$ for each dataset [Figure 7B]. We find that $HAI_{min}$ decreases with ΔVirus Years, demonstrating that it is harder to simultaneously inhibit more diverse viruses. This same tradeoff was seen for monoclonal antibodies (Creanga et al., 2021; Einav et al., 2022), and it suggests that efforts geared towards finding extremely broad, and potentially universal, influenza responses may run into an HAI ceiling.

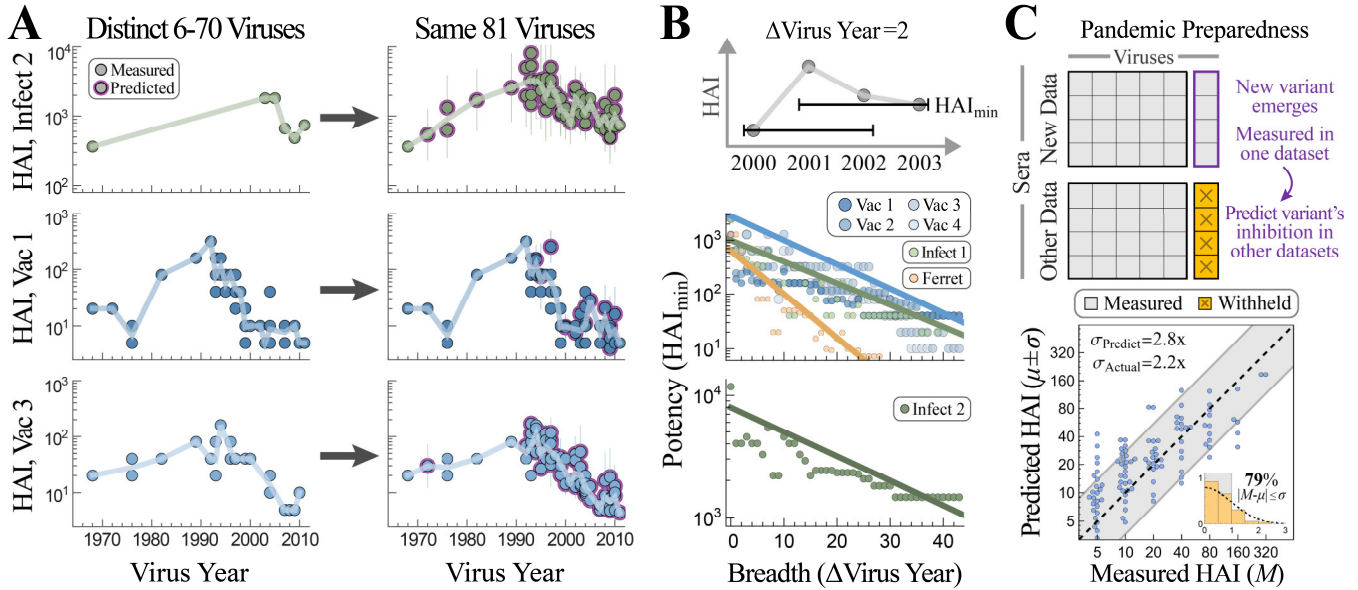

**Figure 7. Applications of cross-study predictions.** (A) We predict HAI titers for 25,000 sera against the same set of 81 viruses, providing high-resolution landscapes that can be directly compared against each other. Representative responses are shown for Dataset$_{Infect,2}$ [*top*, Serum 5130165 in GitHub], Dataset$_{Vac,1}$ [*middle*, Subject 525], and Dataset$_{Vac,3}$ [*bottom*, Subject A028]. (B) Tradeoff between serum breadth and potency, showing that viruses spaced apart in time are harder to simultaneously inhibit. For every study and each possible set of viruses circulating within ΔVirus Years of each other, we calculate the highest HAI$_{min}$ (*i.e.*, a serum exists with HAI titers≥HAI$_{min}$ against the entire set of viruses). (C) *Top*, When a new variant emerges and is measured in a single study, we can predict its titers in all previous studies with ≥5 overlapping viruses. *Bottom*, Example predicting HAI from the most recent virus in the latest vaccine dataset across another vaccine study (Datasets$_{Vac,4 \rightarrow Vac,3}$), demonstrating how information from one context can immediately benefit other surveillance efforts.

**Towards Pandemic Preparedness**

When two studies have high transferability, each serves as a conduit of information for the other, which can greatly amplify the impact of each new experiment. For example, if a new variant emerges and is measured in one study, we can immediately predict this variant's behavior in all other studies with ≥5 overlapping viruses, where the datasets with the highest transferability will yield the most confident predictions. As an example, we consider the more recent virus strain in the latest vaccine dataset (A/Perth/16/2009 from Vaccine Study 4 carried out in 2010, around the time this variant emerged). If we withhold this variant from Vaccine Study 3, we can predict its titers with $\sigma_{Actual}$=2.2-fold error (Figure 7C). This analysis could be applied to new variants emerging today, especially for concerted surveillance efforts, with the transferability and overlap between virus panels quantifying the accuracy of these predictions.

**Matrix Completion via Nuclear Norm Minimization Poorly Predicts Behavior across Studies**

In this final section, we briefly contrast our algorithm against singular value decomposition (SVD) based approaches such as nuclear norm minimization (NNM), which are arguably the simplest and best-studied matrix completion methods. With NNM, missing values are filled by minimizing the sum of singular values of the completed dataset.

To compare our results, we reran our leave-multi-out analysis from Figure 5, simultaneously withholding 124 viruses and predicting their values using an established NNM algorithm from (Einav and Cleary, 2022). The predictions in each table were notably worse, with the Fonville and Vinh datasets exhibiting $\sigma_{Actual}$ between 3.4-5.4x.

Due to two often-neglected features of NNM, we find that our approach significantly outperforms this traditional route of matrix completion in predicting values for a completely withheld virus column. First, one artifact of NNM is that there is an asymmetry when predicting large and small values for the withheld virus. Consider a simple noise-free example where one virus's measurements are proportional to another's, (Virus 2's values)=$m$·(Virus 1's values) [$m$=5 shown in Figure S8A]. Surprisingly, even if provided with one perfect template for these measurements, NNM incorrectly predicts that (Virus 2's values)=(Virus 1's values) for any $m$≥1 [Figure S8B]. This behavior is exacerbated when multiple datasets are combined, emphasizing that NNM can catastrophically fail for very simple examples [Figure S8C,D]. This artifact can be alleviated by first row-centering a dataset before matrix completion, as in Algorithm 1.

Yet even with row-centering, a second artifact of NNM is that large swaths of missing values can skew matrix completion when relationships are (incorrectly) inferred between the missing values. Intuitively, all iterative NNM algorithms must initialize the missing entries (often either with 0 or the row/column means), so that after initialization two viruses with very different behaviors may end up appearing identical across their missing values. For example, suppose we want to predict values for virus $V_0$ from Dataset $X \rightarrow Y$, and that "useful" viruses $V_1$-$V_4$ behave similarly to $V_0$ in Datasets $X$ and $Y$. On the other hand, "useless" viruses $V_5$-$V_8$ are either not measured in Dataset 2 or are measured against complementary sera, and moreover these viruses show very different behavior from $V_0$ in Dataset 1 [Figure S9 shows a concrete example from Fonville]. Ideally, a matrix completion algorithm should ignore $V_5$-$V_8$ (given that they do not match $V_0$ in Dataset 2) and only use $V_1$-$V_4$ to infer $V_0$'s values in Dataset 1. In practice, NNM using $V_0$-$V_8$ results in poor predictions [Figure S9]. This behavior is disastrous for large serological datasets, where there can be >50% missing values when datasets are combined.

Our algorithm was constructed to specifically avoid both of these artifacts. First, we infer each virus's behavior using a decision tree on row-centered data which does not exhibit the asymmetry discussed above. Second, we restrict our analysis to features that have ≥80% observed measurements to ensure that patterns detected are based on measurements rather than on missing data.

As another point of comparison, consider the leave-one-out predictions of the six Vinh viruses using the Fonville datasets. Whereas our algorithm yields tight predictions across the full range of values (Figure S4), NNM led to a nearly flat response with all 25,000 sera incorrectly predicted to be the mean of the measurements (see Figure S11 in (Einav and Cleary, 2022)]. In addition, we utilized an existing SVD based matrix completion method that quantifies the prediction uncertainty for each entry under the assumption that values are randomly missing from a dataset (Chen et al., 2019). Applying this method to the Fonville datasets resulted in predictions whose actual error was >20-fold larger than the estimated error, emphasizing the need for frameworks that specifically handle structured missing data (Hartford et al., 2018).

**Discussion**
By harnessing the wealth of measurements from the past, we can catapult future efforts and design experiments that are far larger in size and scope. Here, we developed an algorithm that leverages patterns in antibody-virus data to predict how a virus measured in one study would behave in another study, without requiring any additional experiments. The resulting high-resolution immunological landscapes can be compared across studies to determine whether vaccination and infection responses fundamentally differ or quantify how accurately ferret data predicts the human antibody response.

A major hurdle in cross-study inferences is reproducibility. Many anecdotes and formal studies have demonstrated that subtle differences in a protocol can lead to vastly different results (Hines et al., 2014; Zacour et al., 2016). Our algorithm enables cross-study predictions in two key ways. First, we determine the transferability between two studies, quantifying how behaviors in one study map to the other (Figure 4B, Figure S1). Using this transferability, we can estimate the uncertainty of each prediction without requiring information about study design or infection history. Second, we account for systematic shifts between datasets by developing a random forest algorithm using row-centered data (Methods). The resulting predictions were sufficiently robust to infer measurements between the Fonville and Vinh studies, even though they utilized different assays, had different dynamic ranges, and used markedly different virus panels (Fonville et al., 2014; Vinh et al., 2021). After validating our predictions on available data, we expanded the Vinh panel to include 75 new viruses with $2 \cdot 10^6$ new measurements, increasing the original dataset by >10x. Similar expansions can and should be carried out for future studies.

Our results suggest that instead of thinking about each serum sample as being entirely unique, large collections of sera often exhibit surprisingly similar inhibition profiles. For example, the 1,200 sera in the Fonville datasets predicted the behavior of the 25,000 Vinh sera with ≤2.5-fold error on average, demonstrating that these Vinh sera were *at least* 20-fold degenerate. As studies continue to probe the diverse pool of sera worldwide, their transferability will determine when new sera exhibit fundamentally different behavior.

We use the transferability between two datasets to succinctly quantify their similarity. It is rarely clear *a priori* when two datasets can inform one another – will differences in age, geographic location, or infection history between individuals fundamentally change how they inhibit viruses (Lewnard and Cobey, 2018; Henry et al., 2019; Dugan et al., 2020)? Transferability directly addresses these questions, with high transferability indicating that information can be readily exchanged between two datasets, enabling new modes of pandemic preparedness (Figure 7C). Low transferability can also be used to probe the antibody response; for example, low transferability between binding and HAI (or neutralization) data on the same group of sera indicates that inhibition (or neutralization) is not just mediated through binding. In this way, functional data on the antibody response informs the underlying mechanism at work.

In this work, we found surprisingly large transferability between human infection and vaccination studies. For example, vaccine studies from 1997/1998 ($Dataset_{Vac,1/2}$) were highly informed by the Vinh infection study from 2009-2015 ($Dataset_{Infect,2}$), even though none of the Vinh participants had ever been vaccinated. Conversely, both infection studies we analyzed were highly informed by vaccine studies (*e.g.*, $Dataset_{Infect,1}$ was most informed by $Datasets_{Vac,3/4}$). These results demonstrate that even studies with very different experimental setups can be combined to predict new virus behavior.

With the transferability relations, we expanded the Fonville and Vinh datasets against 81 H3N2 viruses. Using the age of each subject, these landscapes can quantify how immune imprinting throughout childhood shapes the subsequent antibody response (Vinh et al., 2021). In addition, given the growing interest in universal influenza vaccines capable of inhibiting diverse variants, these high-resolution responses will power studies examining the breadth of the antibody response not just forward in time against newly emerging variants, but also backwards in time to assess how rapidly immunity decays (Carter et al., 2016; Boyoglu-Barnum et al., 2021; Fox et al., 2022). We found that serum potency (the minimum HAI titer against a set of viruses) decreases for more distinct viruses [Figure 7B], as was shown for monoclonal antibodies (Creanga et al., 2021; Einav and Cleary, 2022). Therefore, groups searching for broad and potent antibody responses must balance these two opposing traits. For example, a specific HAI target (*e.g.*, responses with titers≥80 against multiple variants) may only be possible for viruses spanning 1-2 decades.

Our framework inspires new principles of data acquisition, where future studies can save time and effort by choosing smaller virus panels that can be subsequently expanded (Figure 6A). One powerful approach is to perform experiments in stages, interspersing matrix completion to inform next steps. For example, a new study seeking to measure serum inhibition against 100 viruses should first measure 5 viruses that are widely spaced out in time. After computing the transferability of the new dataset using these 5 viruses, the values±errors of the full panel can be estimated. Importantly, we can then predict which additional virus would maximally decrease the prediction errors. Each additional measurement serves as a test for the predictions, and experiments can stop once enough measurements fall sufficiently close to the model predictions.

Matrix completion draws upon the collective information gained in multiple studies to infer how an antibody response would inhibit a broad array of viruses. Such interactions underpin diverse research efforts ranging from viral surveillance (Morris et al., 2018) to characterizing the composition of antibodies within serum (Georgiev et al., 2013; Fonville et al., 2014; Lee et al., 2019a; Einav et al., 2022) to predicting future antibody-virus coevolution (Sheng and Wang, 2021; Marchi et al., 2021). Although we focused on influenza HAI data, this method can be readily applied to other inherently low-dimensional datasets, both in and out of immunology. In the context of antibody-virus interactions, this approach not only massively extends current datasets, but also provides a level playing field where antibody responses from different studies can be directly compared using the same set of viruses. This shift in perspective expands the scope and utility of each measurement, enabling future studies to always build on top of previous results.

**Methods**
*Availability of Code and Results*
The code to perform matrix completion in *Mathematica* and *R* are included in the associated GitHub (https://github.com/TalEinav/CrossStudyCompletion), which also contains the expanded Fonville and Vinh datasets shown in Figure S5.

*Datasets Analyzed*
Information about the Fonville and Vinh datasets (type of study, year conducted, and geographic location) is provided in Table 1. The number of sera, viruses, and missing measurements in each dataset is listed below the schematics in Figure 4. Every serum was unique, appearing in a single study. All Fonville viruses appeared in at least 2 studies, enabling us to entirely remove a virus from one dataset and infer its behavior from another dataset.

Although the Vinh dataset contained H1N1 and H3N2 viruses, we only considered the 6 H3N2 strains (since Fonville only contained H3N2 viruses). 4/6 of the Vinh viruses (H3N2 A/Wyoming/3/2003, A/Wisconsin/67/2005, A/Brisbane/10/2007, and A/Victoria/361/2011) were found in the Fonville datasets. The remaining 2 viruses were not in Fonville, and hence we associated each of them with the Fonville virus that had the most similar HA sequence (Vinh virus A/Aichi/2/1968→Fonville virus A/Bilthoven/16190/1968; Vinh virus A/Victoria/210/2009→Fonville virus A/Hanoi/EL201/2009). While such substitutions may increase prediction error (which can be gauged through leave-one-out analysis), they also vastly increase the number of possible cross-study predictions, since two virus panels only need to have homologous viruses rather than exact matches.

| Influenza Dataset | Organism | Type | Year Conducted | Geographic Location | Source of Data |
|---|---|---|---|---|---|
| $Dataset_{Vac,1}$ | Human | Vaccination | 1997 | Parkville, Australia | Fonville 2014, Table S5 |
| $Dataset_{Vac,2}$ | Human | Vaccination | 1998 | Parkville, Australia | Fonville 2014, Table S6 |
| $Dataset_{Vac,3}$ | Human | Vaccination | 2009 | Parkville, Australia | Fonville 2014, Table S13 |
| $Dataset_{Vac,4}$ | Human | Vaccination | 2010 | Parkville, Australia | Fonville 2014, Table S14 |
| $Dataset_{Infect,1}$[b] | Human | Infection | 2007-2012 | Ha Nam, Vietnam | Fonville 2014, Table S3 |
| $Dataset_{Infect,2}$[b] | Human | Infection | 2009-2015 | Ha Nam, Vietnam | Vinh 2021 Supplement |
| $Dataset_{Ferret}$[a] | Ferret | Infection | N/A | N/A | Fonville 2014, Table S1 |

**Table 1. Datasets analyzed in this work.** Information about the type of study as well as the year and geographic location from which the antibody responses were collected.
[a] Infected influenza-naive ferrets with a single virus and measured their serum against a panel of viruses.
[b] Over multiple years, participants reported influenza-like illnesses and got PCR tested. Serum samples were collected from all participants once each year.

*Matrix Completion on $\log_{10}$(HAI titers)*
The hemagglutination inhibition (HAI) assay quantifies how potently an antibody or serum inhibits the ability of a virus to bind red blood cells. The value (or titer) for each antibody-virus pair corresponds to the maximum dilution at which an antibody inhibits this interaction, so that larger values represent a more potent antibody. This assay is traditionally done using a series of dilutions, so that the HAI titers can equal 10, 20, 40…

As in previous studies, all analysis was done on $\log_{10}$(HAI titers) because experimental measurements span orders of magnitude, and we did not want to bias the predictions towards the largest values (Einav and Cleary, 2022). Thus, when computing the distribution of errors (histogram in Figures 3-5), each of $M$, $\mu$, and $\sigma$ are computed in $\log_{10}$. The only exception is that when presenting the numeric values of a prediction or its error, we did so in un-logged units so the value could be readily compared to experiments. An un-logged value is simply exponentiated by 10 (*i.e.*, $\sigma_{Predict,log10}$=0.3 for $log_{10}$ titers corresponds to an error of $\sigma_{Predict}$=$10^{0.3}$=2-fold, with "fold" indicating an un-logged number). The following sections always refer to $M$, $\mu$, and $\sigma$ in $log_{10}$ units.

In the Fonville dataset, we replaced lower or upper bounds by their next 2-fold increment ("<10"→5 and "≥1280"→2560). The Vinh dataset did not include any explicit bounded measurements, although their HAI titers were clipped to lie between 10-1810, as can be seen by plotting the values of any two viruses across all sera. Hence, the Vinh predictions in Figure 4 (Dataset$_{Infect,2}$) contains multiple points on the left and right edges of the plot.

*Error of the Hemagglutination (HAI) Assay*
In the Fonville 2014 study, analysis of repeated HAI measurements showed that the inherent error of the assay is log-normally-distributed with standard deviation $\sigma_{Inherent}$≈2-fold. This is shown by Figure S8B in (Fonville et al., 2014) [neglecting the stack of not-determined measurements outside the dynamic range of the assay], where 40% of repeats had the same HAI value, 50% had a 2-fold discrepancy, and 10% had a 4-fold discrepancy.

*Using Decision Trees to Quantify the Relationships between Viruses*
Decision trees are a simple, easily-interpretable, and well-studied form of machine learning. An advantage of decision trees is that they are very fast to train, with implementations in many programming languages. The predictions from decision trees are made even more robust using a random forest approach, which in our case involved averaging the results from 5 top trees.

As described in Algorithm 1, we trained regression trees that take as input the row-centered $log_{10}$(HAI titers) from viruses $V_1$-$V_5$ to predict another virus $V_0$. These trees can then be applied in another dataset to predict $V_0$ based on the values of $V_1$-$V_5$.

Row-centering means that if we denote the $log_{10}$(titers) of $V_0$-$V_5$ to be $t_0$-$t_5$ with mean $t_{avg}$, then the decision tree will take as input ($t_1$-$t_{avg}$, $t_2$-$t_{avg}$, $t_3$-$t_{avg}$, $t_4$-$t_{avg}$, $t_5$-$t_{avg}$) and attempt to predict $t_0$-$t_{avg}$. The value $t_{avg}$ (which will be different for each serum) is then added to the output of the decision tree to undo the row-centering. If any of the $t_j$ are missing (including $t_0$ when we withhold $V_0$'s values), we proceed in the same way but compute $t_{avg}$ as the average of the measured values. Row-centering enables the algorithm to handle systematic differences in data. For example, the analysis is independent of which units are chosen for the data (*e.g.*, neutralization measurements in μg/mL or Molar would both be handled the same, since in $log_{10}$ they are offset from each other by a constant factor that will be subtracted during row-centering). If one serum is concentrated by 2x (or any other factor), its titers would all increase by 2x but the relationships between viruses would remain the same; row-centering subtracts this extra concentration factor and yields the same analysis.

When training our decision trees, we allow missing values for $V_1$-$V_5$ (but not $V_0$) [as shown by the schematic in Figure 2B], with these missing values replaced by the most likely value (*i.e.,* mode-finding) given the known values in the training set. When applying a trained decision tree to other datasets, we only predicted a value for $V_0$ when none of $V_1$-$V_5$ were missing (otherwise that decision tree was ignored). If all 5 top trees were ignored due to missing values, then no prediction was made for that virus $V_0$ and serum combination.

**Algorithm 1: Predicting Virus Behavior (Value±Error) across Studies via a Random Forest**

*Input*:
- Dataset-of-interest $D_0$ containing virus-of-interest $V_0$ whose measurements we predict;
- Other datasets $\{D_j\}$, each containing $V_0$ and at least 5 viruses $V_{j,1}$, $V_{j,2}$... that overlap with the $D_0$ virus panel, used to extrapolate virus behavior;
- Antibody responses $A_{j,1}$, $A_{j,2}$... in each dataset $D_j$. When $j \neq 0$, we only consider the antibody responses with non-missing values against $V_0$

*Steps*:
1. For each $D_j$, create $n_{Tree}$=50 decision trees predicting $V_0$ based on $n_{Features}$=5 other viruses and a fraction $f_{Samples}$=3/10 of sera
   - For robust training, we restrict attention to features with ≥80% non-missing values. If fewer than $n_{Features}$ viruses in $D_j$ satisfy this criterion, do not grow any decision trees for this dataset
   - Bootstrap sample (with replacement) both the viruses and antibody responses
   - Data is analyzed in $\log_{10}$ and row-centered on the features (*i.e.*, for each antibody response in both the training set $D_j$ and testing set $D_0$, subtract the mean of the $\log_{10}$[titers] for the $n_{Features}$ viruses using all non-missing measurements). This row-centering accounts for systematic shifts between datasets. Once the decision tree makes its predictions, this row-centering is then undone (by adding this serum-dependent mean)
   - Compute the cross-validation RMSE ($\sigma_{Training}$) of each tree against the unused fraction 1-$f_{Samples}$ of samples in $D_j$
2. Use the $n_{BestTrees}$=5 decisions trees with the lowest $\sigma_{Training}$ to predict the (un-row-centered) values of $V_0$ in $D_0$
   - Only make predictions for antibody responses in $D_0$ where *all* $n_{Features}$ are non-missing. If any feature is missing, this tree makes no prediction for this antibody response
   - For each antibody response, we predict $\mu_j \pm \sigma_j$
     - $\mu_j$=(mean value for the $n_{BestTrees}$ predictions)
     - $\sigma_j = f_{D_j \to D_0}$ (mean $\sigma_{Training}$ for the $n_{BestTrees}$ trees), where the transferability $f_{D_j \to D_0}$ is computed by predicting $V_{j,1}$, $V_{j,2}$... in $D_0$ using $D_j$ (see Algorithm 2)
3. Combine predictions for $V_0$ in $D_0$ using all other datasets $\{D_j\}$ as $\frac{\Sigma_j \, (\mu_j/\sigma_j^2)}{\Sigma_j \, (1/\sigma_j^2)} \pm \frac{1}{[\Sigma_j(1/\sigma_j^2)]^{1/2}}$

*Predicting the Behavior of a New Virus*

As described in Algorithm 1, the values for $V_0$ predicted from Dataset $D_j \to D_0$ is based on the top 5 decision trees that predict $V_0$ in $D_j$ with the lowest $\sigma_{Training}$. The value of $V_0$ against any serum is given by the average value of the top 5 decision trees, while its error is given by the estimated error $\sigma_{Predict} = f_{D_j \to D_0}(\sigma_{Training})$ of these top 5 trees, where $f_{D_j \to D_0}$ represents the transferability map [described in the next section]. Thus, every prediction of $V_0$ in $D_0$ will have the same $\sigma_{Predict}$, unless some of the top 5 trees cannot cast a vote because their required input titers are missing (in which case the value±error is computed as the average of the top 5 trees that can vote). In practice, the estimated error for $V_0$ in $D_0$ is overwhelmingly the same across all sera, as seen in Figure 3 where the individual error of each measurement is shown via error bars.

The estimated error $\sigma_{Predict}$ and true error $\sigma_{Actual}$ in Figures 4 and 5 were computed using all data. When scatter plots contained too many data points to show with appreciable resolution, we subsampled each distribution evenly across its predicted value to maintain its shape. We did not display the small fraction of measurements with HAI titers≥640 in order to better show the portions of the plots with the most points; however, error statistics were computed using all data.

*Transferability Maps between Datasets*
A key component of our analysis is to quantify how the error of a decision tree trained in dataset $D_j$ translates into this tree's error in dataset $D_0$. Importantly, when predicting the behavior of a virus $V_0$ in $D_0$, we *cannot* access $V_0$'s values and hence cannot directly compute the actual error of this tree.

To solve this problem, we temporarily ignore $V_0$ and apply Algorithm 1 to predict the titers of viruses measured in both $D_0$ and $D_j$. Since we can access the values of these viruses in both datasets, we can directly compare their $\sigma_{Training}$ in $D_j$ against $\sigma_{Actual}$ in $D_0$. We did not know *a priori* what the relationship would be between these two quantities, and surprisingly, it turned out to be well-characterized by a simple linear relationship $f_{D_j \to D_0}$ (Figure S1, blue lines). As described in the following paragraph, these relations represent an upper bound (*not* a best fit) through the ($\sigma_{Training}$, $\sigma_{Actual}$) points, so that our estimated error $\sigma_{Predict} \equiv f_{D_j \to D_0}(\sigma_{Training})$ obeys $\sigma_{Predict} \geq \sigma_{Actual}$; thus, when we estimate a small $\sigma_{Predict}$ we expect $\sigma_{Actual}$ to be small, whereas a large $\sigma_{Predict}$ implies that in the worst-case we have a large $\sigma_{Actual}$, although we may also be pleasantly surprised with a smaller actual error.

Following Algorithm 2, we obtain the best-fit line to these data (using perpendicular offsets for a more intuitive fit). To account for the scatter about this best-fit line (which arises when some viruses are accurately predicted across two datasets while others are not), we add a vertical shift given by the RMSE of the deviations from the best-fit line, thereby ensuring that in highly-variable cases where some trees have small $\sigma_{Training}$ but large $\sigma_{Actual}$ (*e.g.*, Dataset$_{Ferret}$→Dataset$_{Vac,1}$), we tend to overestimate rather than underestimate the error.

To visualize the transferability maps between every pair of datasets, we construct a chord diagram where the arc connecting Dataset *X* and *Y* represents a double-sided arrow quantifying both the transferability from *X*→*Y* [thickness of the arc at Dataset *Y*] as well as the transferability from *Y*→*X* [thickness of the arc at Dataset *X*] (Figure S3). The width of each arc is equal to $\Delta\theta \equiv (2\pi/17)(\partial f_{D_j \to D_0}/\partial \sigma_{Training})^{-1}$, so that the width is proportional to 1/slope of the transferability best-fit line from Figure S1. We used the factor 17 in the denominator so that the chord diagrams in Figures 4B,5B would form nearly complete circles, and if more studies are added this denominator can be modified (increasing it would shrink all the arcs proportionally). Note that the size of the arcs in Figures 4B and 5B can be directly compared to one another, so that if the arc from *X*→*Y* is wider in one figure, it implies more transferability between these datasets. A chord connects every pair of studies, unless there were fewer than 5 overlapping viruses between the studies (*e.g.*, between Dataset$_{Infect,2}$ and Dataset$_{Vac,3/4}$), in which case the transferability could not be computed.

### Algorithm 2: Computing the Transferability $f_{D_j \to D_0}$ between Datasets

*Input*:
- Datasets $\{D_j\}$ that collectively include the viruses $V_1$, $V_2$… Each virus may be missing from some datasets, but must be included in at least two datasets

*Steps*:
- For each dataset $D_0 \in \{D_j\}$
  - For each virus $V_0$ in $D_0$
    - For every other dataset $D_j$ containing $V_0$
      - Create $n_{Tree}=50$ decision trees predicting $V$ based on $n_{Features}=5$ other viruses, as described in Algorithm 1
      - For each tree, store the following information:
        - $D_0$, $V_0$, and $D_j$ used to construct the tree
        - Viruses used to train the tree
        - RMSE $\sigma_{Training}$ on the $1-f_{Samples}$ samples in $D_j$
        - Predictions of $V_0$'s values in $D_0$
        - True RMSE $\sigma_{Actual}$ of these predictions for $V_0$ in $D_0$
- When predicting $V_0$ using $D_j \to D_0$ in Algorithm 1, we compute the relationship $f_{D_j \to D_0}$ between $\sigma_{Training}$ and $\sigma_{Actual}$ by predicting the other viruses $V_1, V_2…V_n$ that overlap between $D_j$ and $D_0$ (making sure to only use decision trees that exclude the withheld $V_0$)
  - From the forest of decision trees above, find the top 10 trees for each virus predicted between $D_j \to D_0$ and plot $\sigma_{Training}$ vs $\sigma_{Actual}$ for all trees [see Figure S1]
  - Find the best-fit line using perpendicular offsets, $y=ax+b$ where $x=\sigma_{Training}$ and $y=\sigma_{Actual}$. Since there is scatter about this best-fit line, and because it is better to overestimate rather than underestimate error, we add a correction factor $c$=(RMSE between the $\sigma_{Actual}$ and $ax+b$). Lastly, we expect that a decision tree's error in another dataset will always be at least as large as its error on the training set ($\sigma_{Actual} \geq \sigma_{Training}$), and hence we define $f_{D_j \to D_0} = \max(a\sigma_{Training}+b+c, \sigma_{Training})$. This max term is important in a few cases where $f_{D_j \to D_0}$ has a very steep slope but some decision trees have small $\sigma_{Training}$
  - Datasets with high transferability will have $f_{D_j \to D_0}(\sigma_{Training}) \approx \sigma_{Training}$, meaning that viruses can be removed from $D_0$ and accurately inferred from $D_j$. In contrast, two datasets with low transferability will have a nearly vertical line, $\partial f_{D_j \to D_0}/\partial \sigma_{Training} \gg 1$, signifying that viruses will be poorly predicted between these studies
  - In the chord diagrams (Figures 4B,5B), the width of the arc between Dataset $D_j$ and $D_0$ is proportional to $(\partial f_{D_j \to D_0}/\partial \sigma_{Training})^{-1}$

---

The transferability in Figure 4B and Figure S1 represents all antibody-virus data, which is slightly different from the transferability maps we use when predicting values for virus $V_0$ in dataset $D_0$. When withholding a virus, we made sure to remove from Figure S1 all trees that use this virus as a feature. Although this can slightly change the best-fit line, in practice the difference is very minor. However, when withholding multiple viruses in our leave-multi-out analysis, the number of datapoints in Figure S1 substantially decreased, and to counter this we trained additional decision trees (as described in the following section).

*Leave-Multi-Out Analysis*
For this analysis, we trained many decision trees using different choices of viruses $V_1$-$V_5$ to predict $V_0$ in different datasets. For leave-one-out analysis, we created 50,000 trees, providing ample relationships between viruses. However, when we withheld 124 viruses during the leave-multi-out analysis, we were careful to not only exclude decision trees predicting one of these withheld viruses (as $V_0$), but to also exclude decision trees using any withheld virus in the feature set (in $V_1$-$V_5$). As a result, only 6,000 trees out of our original forest remained, and this smaller number of trees leads to higher $\sigma_{Predict}$ and $\sigma_{Actual}$ error. Fortunately, this problem is easily countered by growing additional trees that specifically avoid the withheld viruses. Once these extra trees were grown, we applied Algorithm 1 as before.

To find a minimal virus panel, we randomly choose one of the 317 virus-study pairs from the Fonville/Vinh datasets, adding it to the list of withheld viruses provided that all withheld entries could be predicted with ≤4-fold error. We note that given a forest of decision trees, it is extremely fast to test whether a set of viruses all have $\sigma_{Predict}$≤4-fold. However, as described above, as more viruses are withheld, our forest is trimmed which leads to poorer estimations of $\sigma_{Predict}$. Hence we worked in stages, interspersing pruning the list of viruses with growing more decision trees.

Our procedure to find a minimal virus panel proceeded in three steps:
- *Step 1*: *Choose Vinh viruses to withhold, and then choose viruses from the Fonville human studies.* Because there are only 6 Vinh viruses, and removing any one of them from the Fonville datasets could preclude making any Vinh predictions, we first withheld 2 Vinh viruses. We then started withholding viruses from the human datasets (Dataset$_{Vac,1-4}$ and Dataset$_{Infect,1}$) where we had the most decision trees.
- *Step 2*: *Create an additional random forest for the Fonville ferret dataset (Dataset$_{Ferret}$).* Make sure to only use the non-withheld viruses from the other datasets as features (since any trees using those withheld viruses will be discarded). As described above, this additional forest quantifies the relationships between the remaining non-withheld viruses with higher resolution, providing more accurate estimations of $\sigma_{Predict}$. With this forest, choose additional viruses from the ferret dataset to withhold.
- *Step 3*: *Create additional random forests for the Fonville human datasets.* Use the improved resolution provided by these new forests to determine if any of the previously withheld viruses now have $\sigma_{Predict}$>4-fold and remove them. Finally, use the additional high-resolution forests to search for additional Fonville viruses to withhold.

*Extending Virus Panels*
To extend the Fonville and Vinh datasets, we grew another forest of decision trees. Unlike in our leave-one-out analysis, the two key differences with this forest were that none of the data were withheld and that the feature set when expanding dataset $D_0$ was restricted to only the viruses within $D_0$. For example, to expand the Vinh dataset and predict one of the 81-6=75 Fonville viruses $V_0$ (excluding the 6 viruses already in the Vinh data), we only searched for relationships between the six Vinh viruses and $V_0$ across the Fonville datasets. (Any decision trees using other viruses could not be applied to the Vinh dataset, which only contains these 6 viruses.)

Once these additional trees were grown, we could predict the behavior of all 81 Fonville viruses against nearly every serum analyzed in the Fonville or Vinh datasets. The exceptions were sera such as those shown in the middle and bottom of Datasets$_{Vac,1/2}$ – these sera were measured against few viruses, and hence we found no relationship between their available measurements in our random forest. The expanded virus panels are available in the GitHub repository associated with this paper.

With the expanded panels, we computed the tradeoff between serum potency and breadth as follows. For every range of ΔVirus Years (for example, ΔVirus Years=2 is shown schematically at the top of Figure 7B), we considered every interval within our dataset (1968-1970, 1969-1971, …, 2009-2011). Note that we only considered intervals where at least one virus was at the earliest and latest year to ensure that the virus set spanned this full range (*e.g.*, we would not consider the interval 1971-1973 since we had no viruses in 1971 or 1973). For each interval, we took all of the 81 viruses that circulated during that interval, and for each serum we computed the weakest response (minimum titer) against any virus in this set. Figure 7B plots the largest minimum titer (HAI$_{min}$) found in each table for any interval of ΔVirus Years, demonstrating that serum potency decreases when inhibiting viruses spanning a broader range of time.

## Acknowledgements

We would like to thank Andrew Butler, Ching-Ho Chang, Bernadeta Dadonaite, David Donoho, and Katelyn Gostic for their input on this manuscript. Tal Einav is a Damon Runyon Fellow supported by the Damon Runyon Cancer Research Foundation (DRQ 01-20). Rong Ma is supported by Professor David Donoho at Stanford University.

## Author Contributions

T.E. and R.M. conducted the research and wrote the paper.

## Competing Interests

The authors declare no competing interests.

# SI Figures

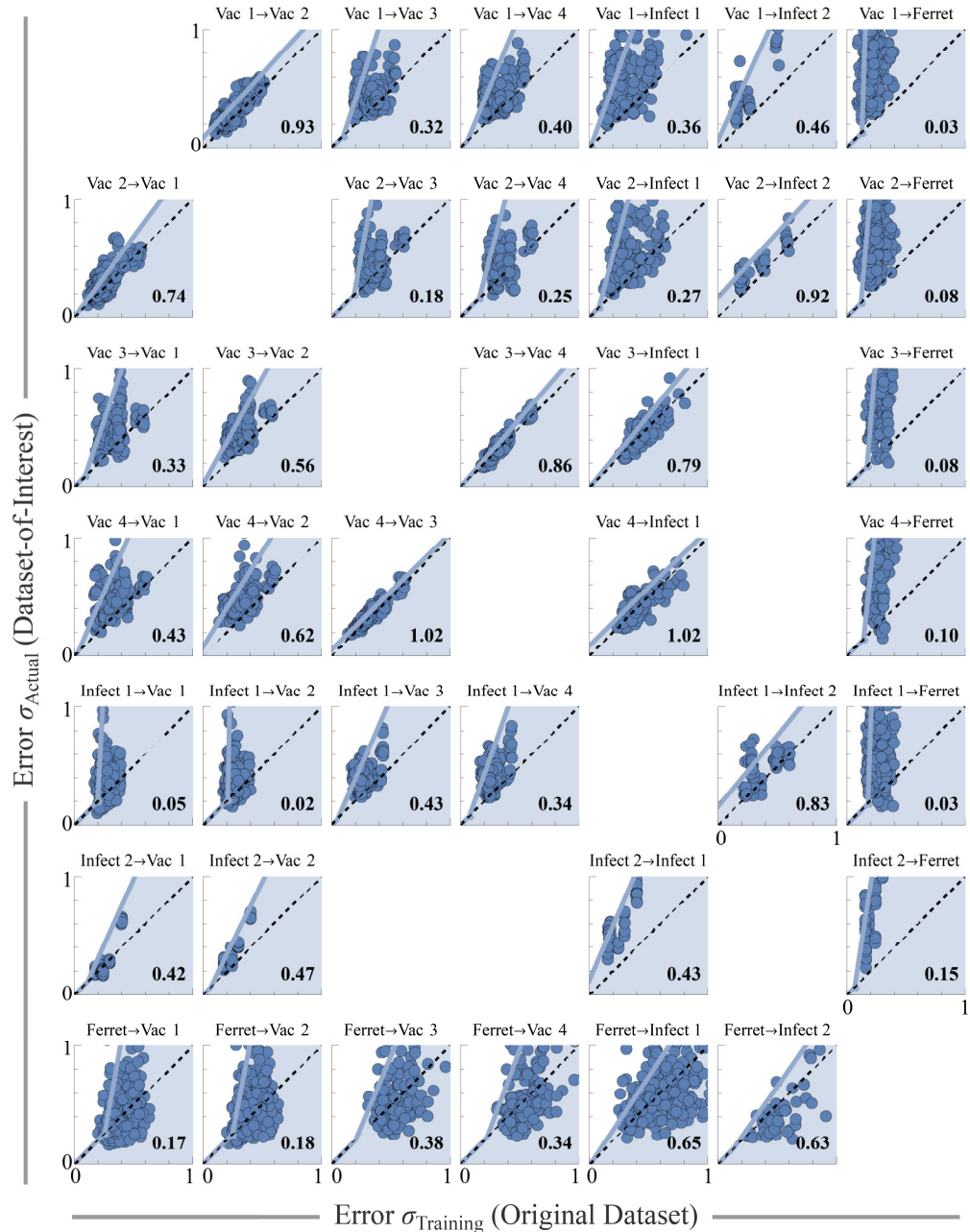

**Figure S1. Transferability of behavior between studies examined in this work.** Each plot quantifies the transferability relation $f_{j\to k}$ of virus behavior between Dataset $j$ and Dataset $k$; the relation $f_{j\to k}$ represents an upper bound (*not* a best fit-line), with the majority of points expected to lie within the shaded region. Each point represents a decision tree trained on 30% of samples in Dataset $j$, with its cross-validation RMSE $\sigma_{Training}$ computed on $\log_{10}$(titers) against the remaining 70% of samples [*x*-axis]. This tree was then applied to Dataset $k$, with RMSE $\sigma_{Actual}$ [*y*-axis]. Every possible virus (measured in both Dataset $j$ and Dataset $k$) was withheld and predicted, and the plotted points represent the 5 decision trees with the lowest $\sigma_{Training}$ (or the top 10 trees if there are fewer than 300 points in the plot to ensure sufficient sampling). The best-fit perpendicular line $f_\perp$ was fit to the resulting points, and to account for variability (and to overestimate rather than underestimate error) we add to this line the constant $f_{RMSE}$ (the RMSE of the vertical deviations between $f_\perp$ and each point). Lastly, because error should increase when extrapolating the predictions to a new dataset ($\sigma_{Training} \leq \sigma_{Actual}$), and because some of the lines are nearly vertical, we enforce that $f_{j\to k}$ lies above $y=x$ by defining $f_{j\to k}=\max(f_\perp+f_{RMSE}, \sigma_{Training})$. The only plots that are not shown are the diagonal entries (we do not need self-transferability) and Vac 3/4 and Infect 2 (these datasets only have 1 overlapping virus which is not enough to quantify transferability; hence no predictions were made between these datasets). The numbers at the bottom-right of each plot show the transferability, 1/(slope of $f_\perp+f_{RMSE}$).

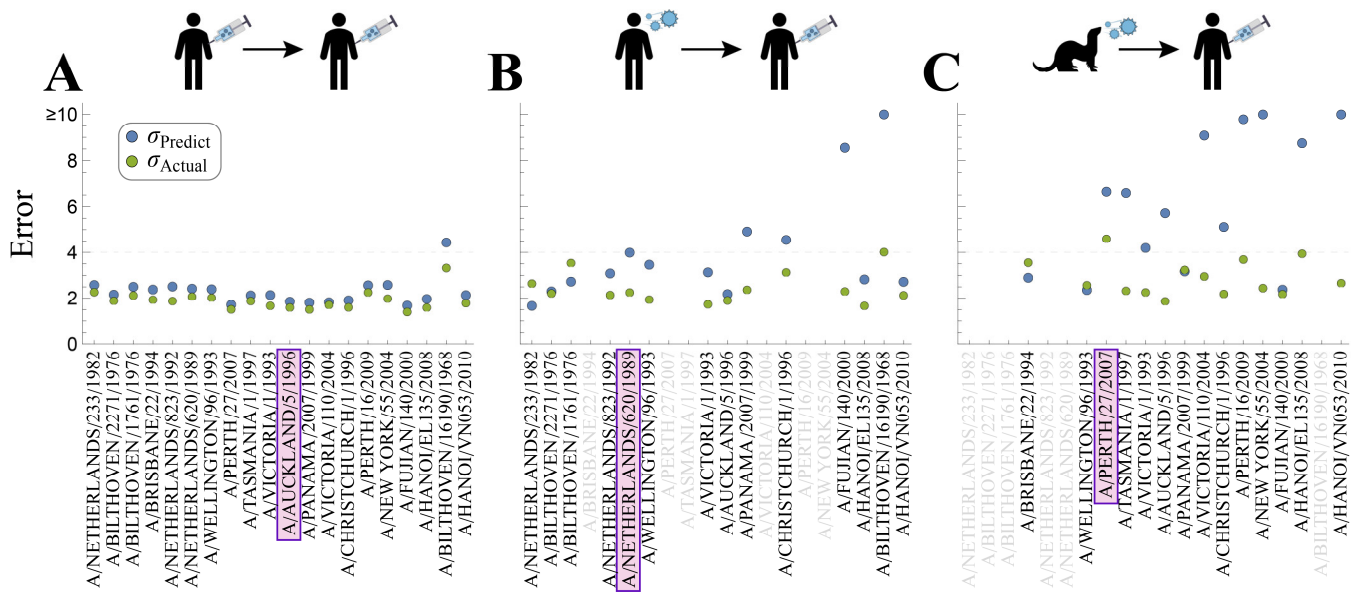

**Figure S2. Predicting each virus in Dataset$_{Vac,4}$ using one other dataset.** We withhold one virus in Dataset$_{Vac,4}$ (*x*-axis) and predict it using (A) the human vaccination study [Dataset$_{Vac,3}$], (B) the human infection study [Dataset$_{Infect,1}$], or (C) the ferret infection study (Dataset$_{Ferret}$). In each case, we show the estimated error ($\sigma_{Predict}$, blue) and the true error ($\sigma_{Actual}$, green). Viruses appear in the same order in each plot, sorted from least-to-greatest average $\sigma_{Predict}$ across the three plots. Grayed-out viruses could not be predicted either because they were absent from a dataset (*e.g.*, Dataset$_{Infect,1}$ did not contain A/Brisbane/22/1994) or because there was insufficient data. The three viruses shown in Figure 3 are boxed in purple.

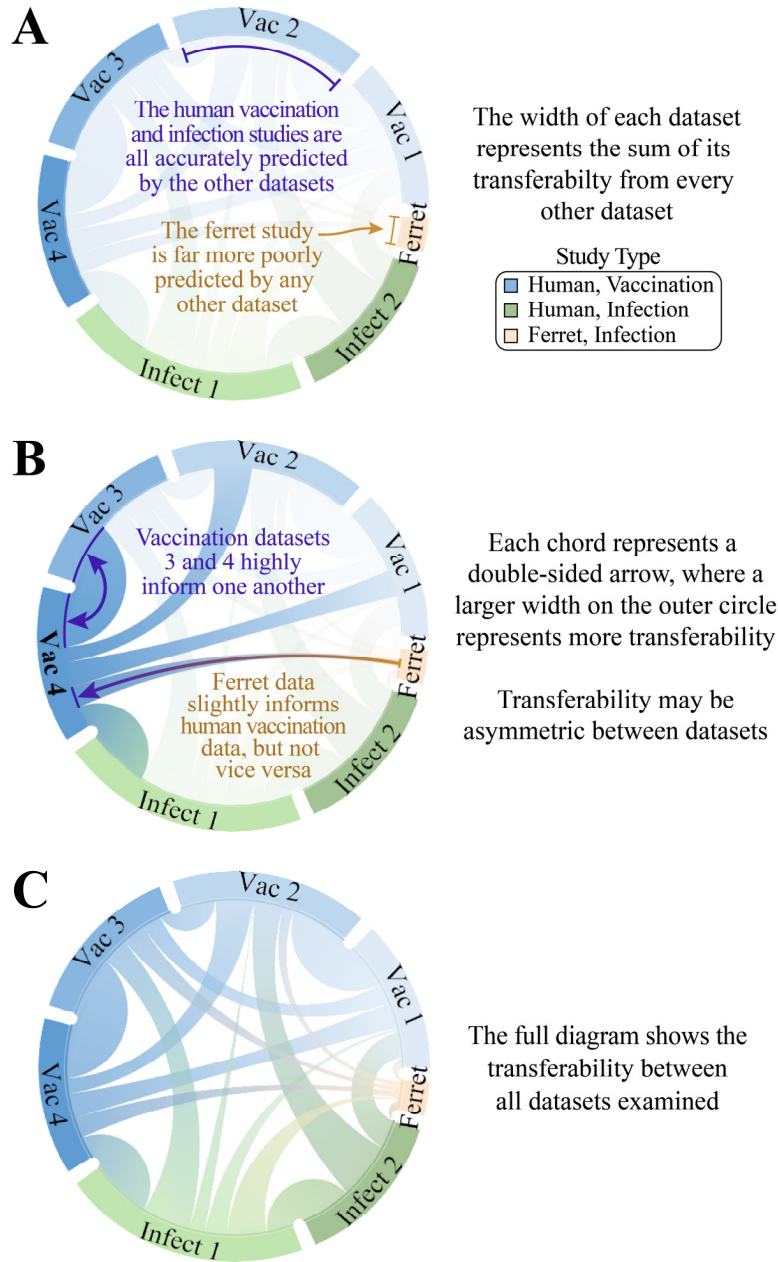

**Figure S3. Explaining the transferability chord diagram.** Figure 4B represents the transferability between the influenza datasets when considering all data. (A) The width of each dataset represents the sum of its transferability from all other datasets. This total width is not directly used (we only use the transferability between each pair of studies), but the smaller total width of the ferret study indicates that all other datasets poorly infers the ferret measurements. (B) A wider arc from Study X→Study Y represents greater transferability. More precisely, transferability equals 1/slope of the linear map in Figure S1, so that studies with near-perfect transferability (slope≈1) will have large width while studies with poor transferability (slope≫1) will have small width. (C) The full diagram from Figure 4B.

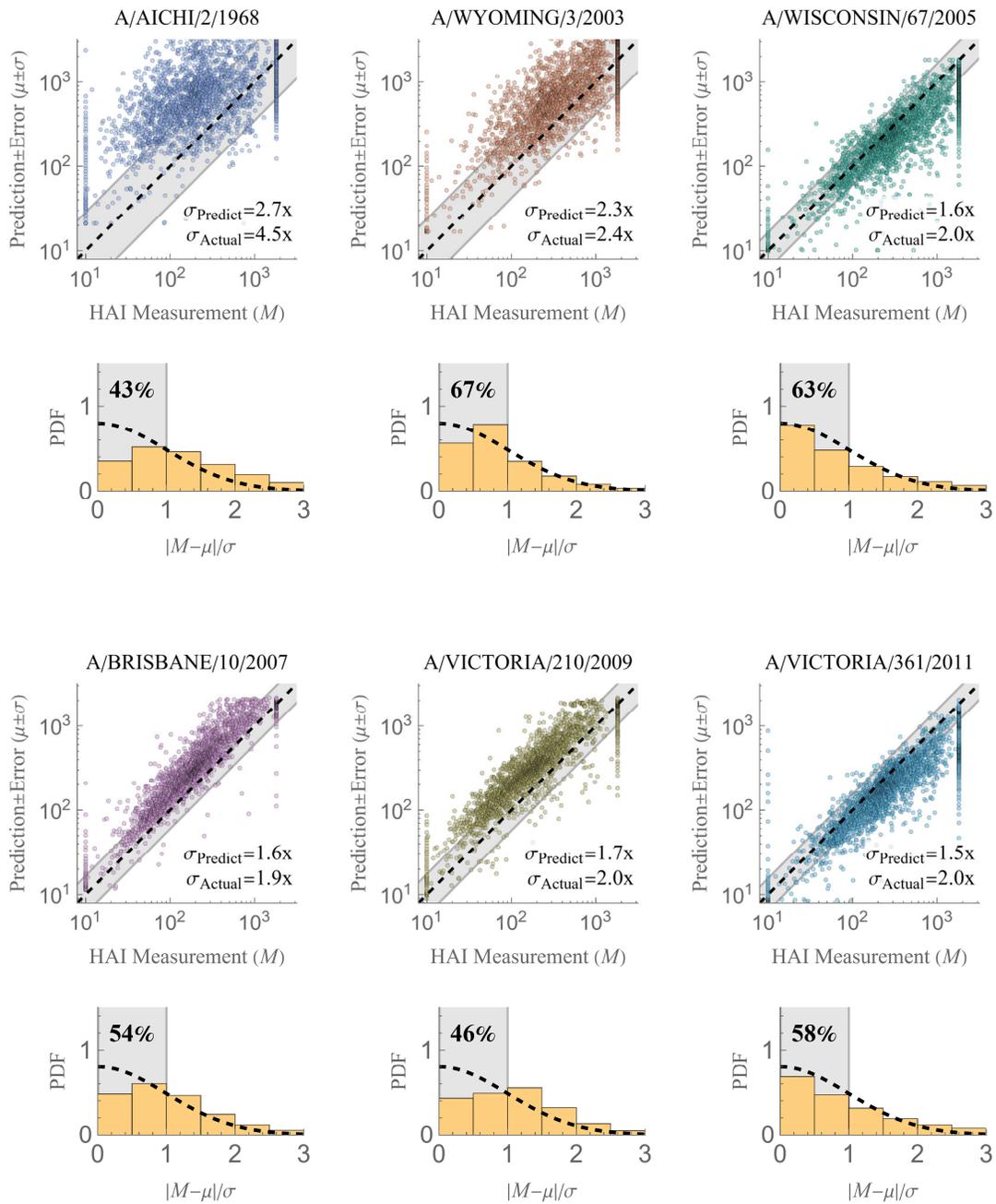

**Figure S4. Individual matrix completions in the Vinh dataset.** Each of the six Vinh viruses were withheld and predicted using the Fonville data. *Scatterplots* show predictions versus measurements. For each virus, the uncertainty of its predictions will be the same for all 25,000 values, and this uncertainty is visualized using the gray bands (showing the fold-error $\sigma_{\text{Predict}}$); the predicted and true errors are also written at the bottom-right of each plot. For clarity, we only show every 10th data point of the 25,000 measurements, but all statistics are computed using the full data. *Histograms* portray the error distribution for the predictions, with the value in the gray region showing the number of predictions within 1σ of the measurement.

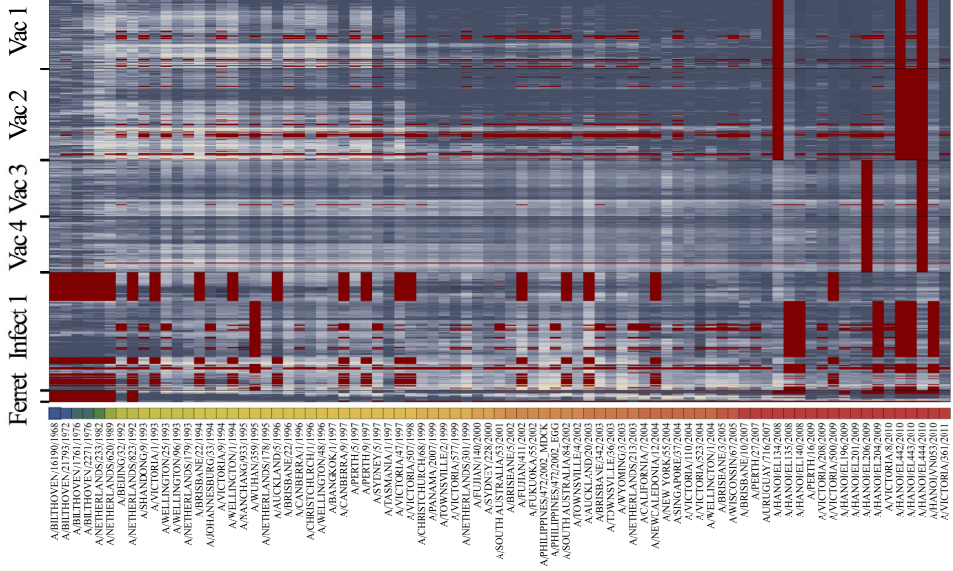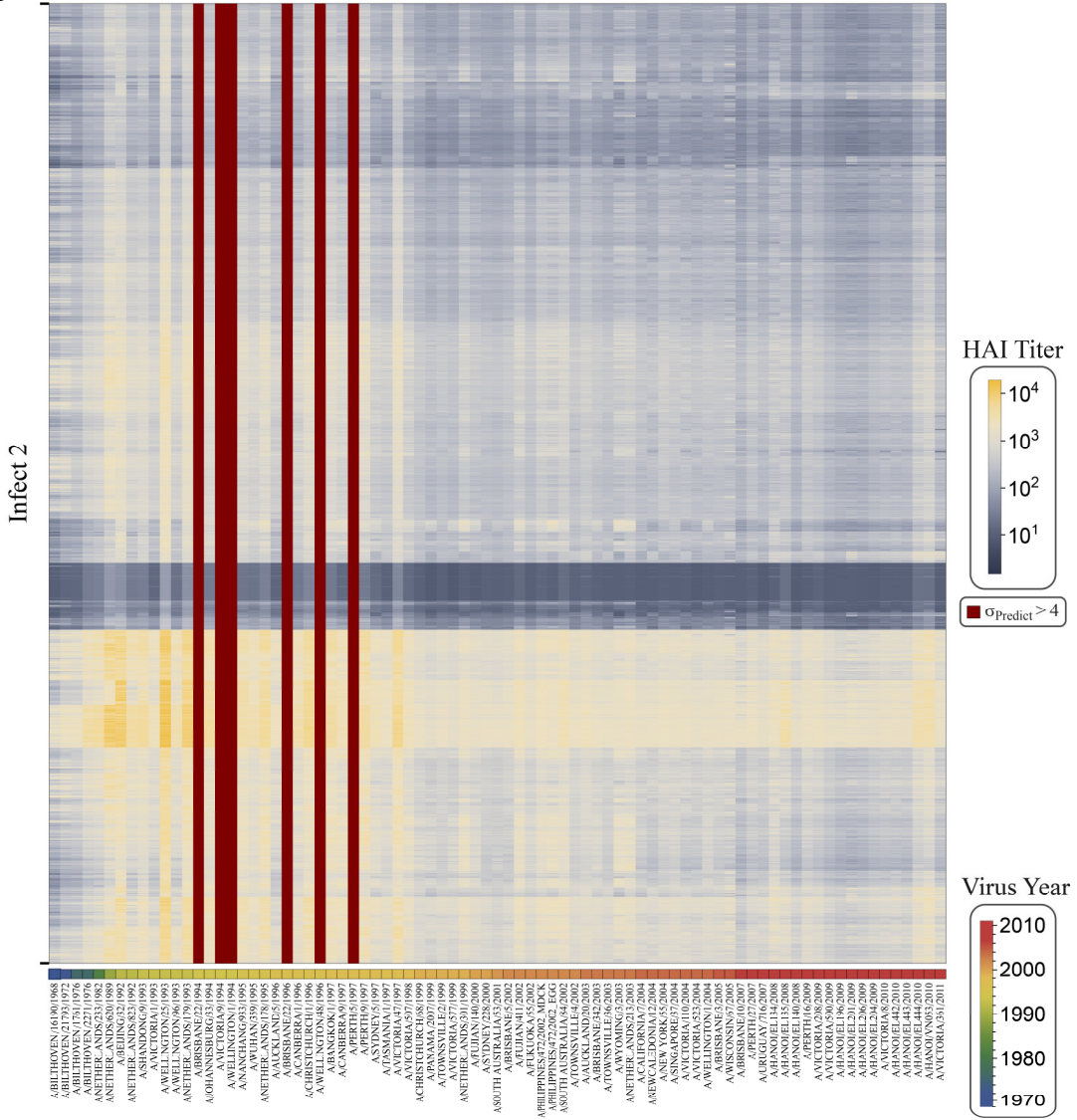

**Figure S5. Expanded HAI titers for all datasets considered in this work.** Using the available measurements, we predicted all antibody-virus interactions in the (A) Fonville and (B) Vinh datasets. In total, we added 32,000 and 1,700,000 new measurements with ≤4-fold error in the Fonville and Vinh datasets, respectively; all other predictions with $\sigma_{Predict}$>4 are shown in dark-red. The sera in each table were clustered based on their Ward similarity function. Viruses are ordered by their year of circulation in both plots, and the color in the bottom row represents a virus's year of circulation. The complete list of measurements and predictions is included in the associated GitHub repository.

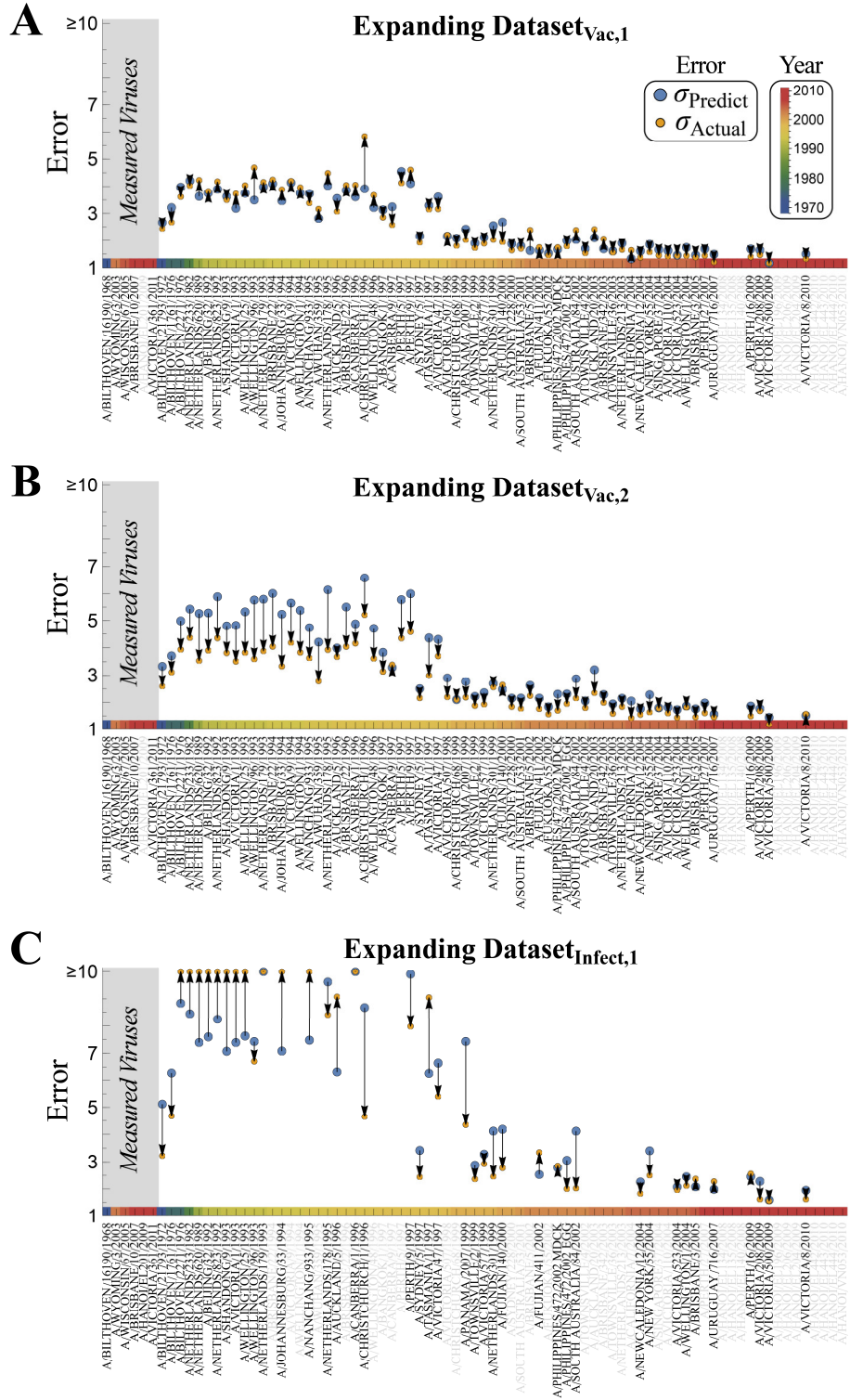

**Figure S6. Extrapolating virus behavior in the Fonville datasets using 6 viruses.** Analogous to Figure 6, we only use values from the six Vinh viruses (or the subset of these viruses present in each Fonville dataset) to predict the behavior of all other viruses. We consider predictions in (A) Dataset$_{Vac,1}$, (B) Dataset$_{Vac,2}$, or (C) Dataset$_{Infect,1}$, which are the three datasets that contribute the most of the Vinh predictions [Figure 4B]. Each plot shows the predicted error [$\sigma_{Predict}$, blue] and actual error [$\sigma_{Actual}$, gold], with a connecting arrow. Viruses in gray could not be predicted either because they were not in the Fonville dataset or there was insufficient data. Viruses from the 1980s and 1990s (which are the furthest away from the 5-6 measured viruses) have the largest error, and this error is slightly overestimated in Dataset$_{Vac,2}$ and underestimated in Dataset$_{Infect,1}$. As explained in the Methods, our framework is constructed so that low $\sigma_{Predict}$ always implies a low $\sigma_{Actual}$ (with $\sigma_{Predict} \approx \sigma_{Actual}$), whereas large $\sigma_{Predict}$ implies less certainty in $\sigma_{Actual}$. A good rule of thumb from these results is to not use values with a predicted error ≥6-fold, since their true error may be even larger; we note that all inferred values in Figure 6C have a predicted error <6-fold.

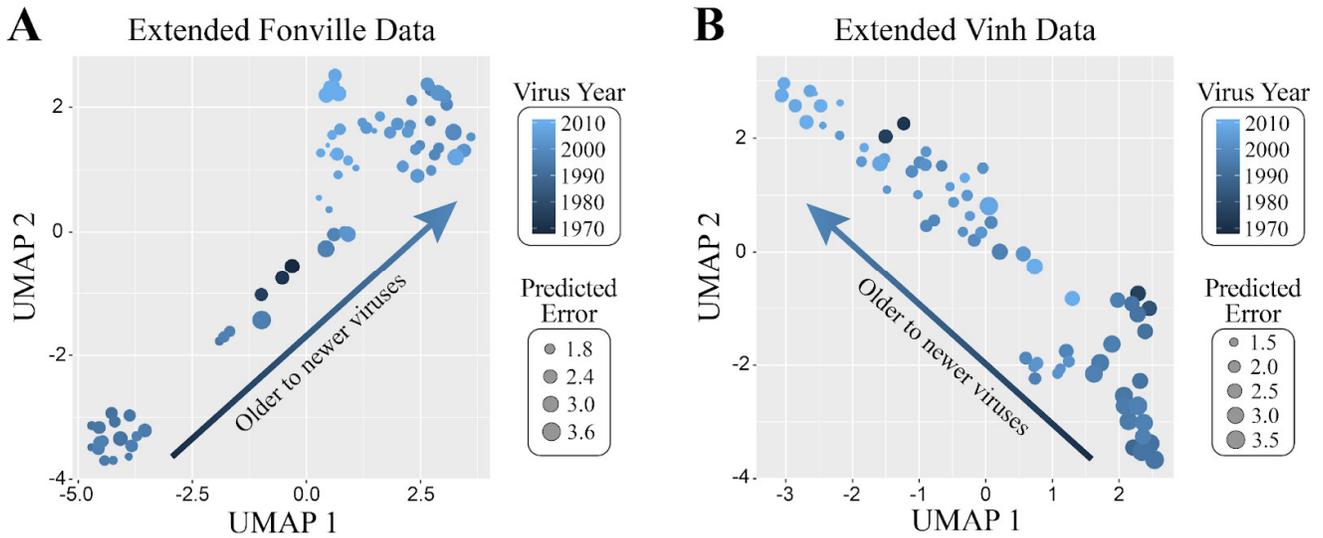

**Figure S7. UMAP embeddings of the expanded antibody-virus datasets.** We applied UMAP to the (A) Fonville and (B) Vinh expanded data, using the log$_{10}$(HAI titers) with $n_{neighbor}$=20 and the default tuning parameters in the *R* package *uwot*. Each data point in the plot corresponds to a specific virus (81 in total), with the size of each point indicating the predicted error of the imputed values, and the shading indicating the year the virus circulated. In both UMAPs, the viruses show a clear temporal pattern moving along a straight line, even though this temporal information was never provided to the algorithm.

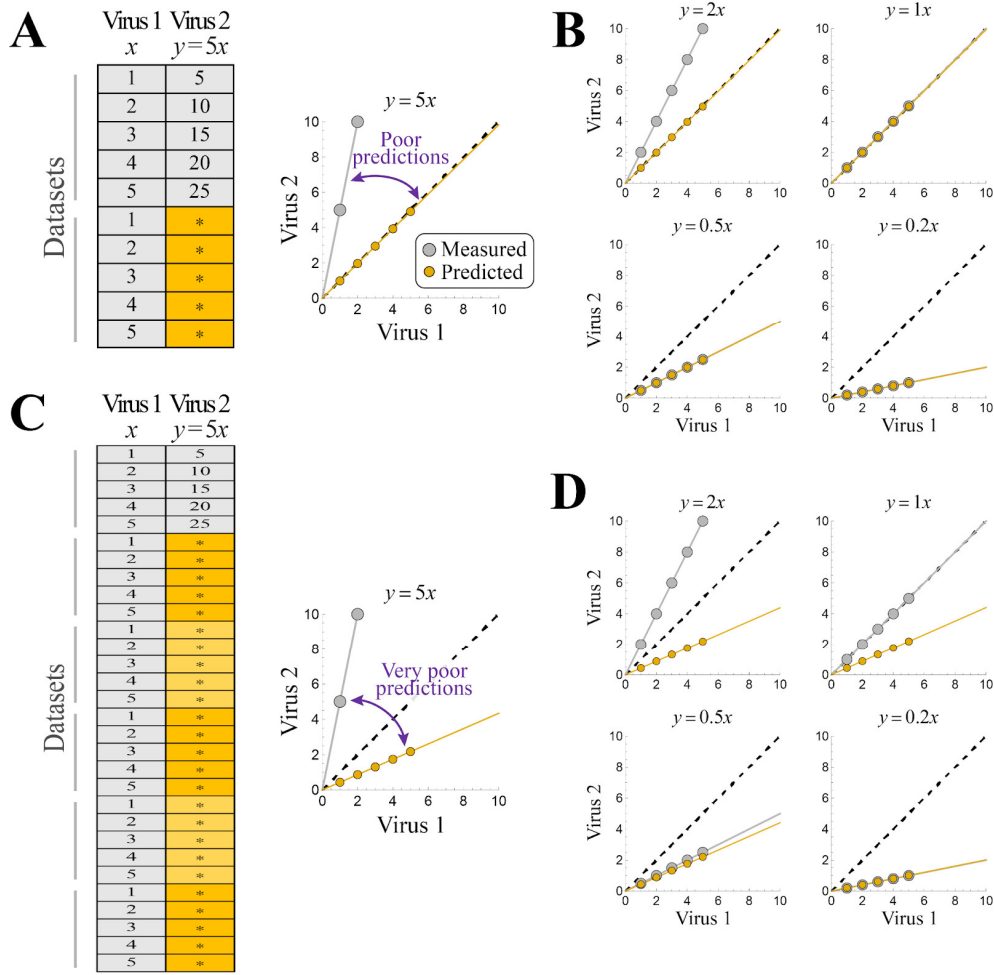

**Figure S8. Nuclear norm minimization can fail in a simple, noise-free setting.** (A) A noise-free toy example where the measurements of two viruses are proportional to each other ($y=5x$) and the input dataset has a perfect template of this relationship, but Virus 2 is nevertheless incorrectly predicted as $y=x$. (B) This problem holds for any relation $y=mx$ where $m>1$, although values of $m \leq 1$ lead to perfect recovery. (C,D) The same setup with $n$ copies of the missing measurements. The problem is now exacerbated, with Virus 2 predicted as $y=n^{-1/2}x$ whenever $m>n^{-1/2}$.

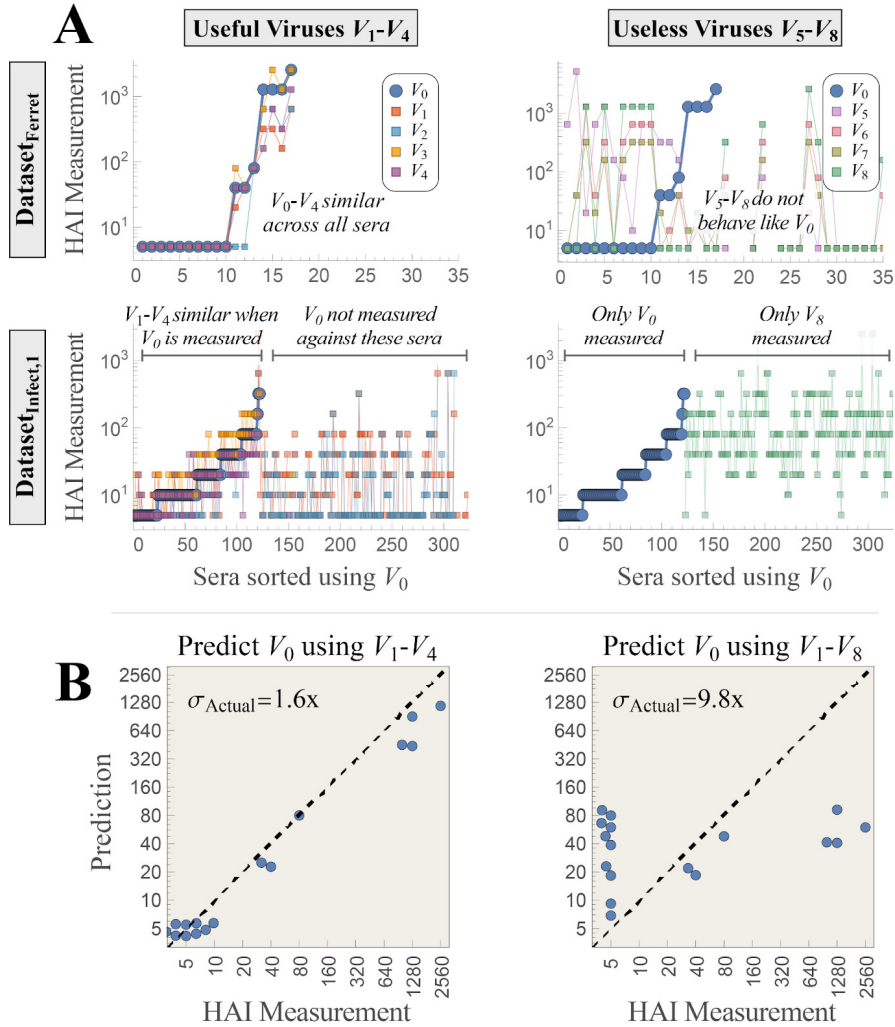

**Figure S9. Nuclear norm minimization may give poor predictions when there are large swaths of missing values.**
Predictions for virus $V_0$ [specified below] from Datasets$_{\text{Infect,1}\to\text{Ferret}}$ are highly accurate when using the "useful" viruses $V_1$-$V_4$ that behave similarly in both studies, but highly inaccurate when adding the additional "useless" viruses $V_5$-$V_8$ that don't behave like $V_0$ in either study. (A) Plot of the titers of the useful and useless viruses in both datasets, with sera sorted according to the HAI titers of $V_0$. Values for $V_1$-$V_4$ closely match those of $V_0$ for all sera in Dataset$_{\text{Ferret}}$ and for all sera where $V_0$ is measured in Dataset$_{\text{Infect,1}}$ (the first 125 sera). In contrast, $V_5$-$V_8$ do not behave like $V_0$ in Dataset$_{\text{Ferret}}$; in Dataset$_{\text{Infect,1}}$ viruses $V_5$-$V_7$ are never measured, and $V_8$ is only measured against sera where $V_0$ was not measured. Hence, $V_5$-$V_8$ should ideally not influence the matrix completion of $V_0$. (B) The resulting predictions vs measurements for $V_0$ only using $V_1$-$V_4$ [left] or using both $V_1$-$V_4$ and $V_5$-$V_8$ [right], with the latter leading to significantly larger error. In the Fonville datasets, these viruses represent $V_0$=VN018/EL204/2009, $V_1$-$V_4$={HN201/2009, HN206/2009, VN019/EL442/2010, VN020/EL443/2010}, $V_5$-$V_8$={A/Singapore/37/2004, A/South Australia/53/2001, A/Sydney/228/2000, A/South Australia/84/2002}.